\documentclass[11pt]{article}
\usepackage[utf8]{inputenc}
\usepackage{bm, commath}
\usepackage{natbib}
\usepackage[hyphens]{url}
\usepackage{caption}
\usepackage{graphicx}
\usepackage{subcaption}
\usepackage{amsmath, amsfonts, amsthm}
\usepackage{float}
\usepackage{booktabs,siunitx}
\usepackage{multirow}
\usepackage{amssymb}
\usepackage{blkarray}
\usepackage{bbm}
\usepackage{multicol}
\usepackage{authblk}
\usepackage[usenames,dvipsnames]{xcolor}
\usepackage{listings}
\usepackage{bbm}
\usepackage{textcomp}
\usepackage[margin=1in]{geometry}
\usepackage{authblk}
\usepackage[ruled]{algorithm2e}
\SetKwInput{KwParam}{Parameter}
\SetAlgoCaptionLayout{centerline}

\usepackage{sectsty}
\usepackage[onehalfspacing]{setspace}
\usepackage{float}
\usepackage{thmtools}
\declaretheorem[style=definition]{example}

\usepackage{tikz}
\usetikzlibrary{positioning,chains,fit,shapes,calc}

\newtheorem{theorem}{Theorem}
\newtheorem{definition}{Definition}
\newtheorem{corollary}{Corollary}

\newtheorem{assumption}{Assumption}
\newtheorem*{assumption*}{Assumption}

\newtheorem{proposition}{Proposition}

\newtheorem{lemma}{Lemma}

\renewcommand\thmcontinues[1]{Continued}

\usepackage{mathtools}

\makeatletter
\renewcommand{\algocf@captiontext}[2]{#1\algocf@typo. \AlCapFnt{}#2} 
\def\@algocf@capt@plain{top}
\renewcommand{\algocf@makecaption}[2]{%
  \addtolength{\hsize}{\algomargin}%
  \sbox\@tempboxa{\algocf@captiontext{#1}{#2}}%
  \ifdim\wd\@tempboxa >\hsize
  \hskip .5\algomargin%
  \parbox[t]{\hsize}{\algocf@captiontext{#1}{#2}}
  \else%
  \global\@minipagefalse%
  \hbox to\hsize{\box\@tempboxa}
  \fi%
  \addtolength{\hsize}{-\algomargin}%
}
\makeatother

\sectionfont{\bfseries\large\sffamily}%
\subsectionfont{\bfseries\sffamily\normalsize}%

\newcommand{\reviewer}[3]{
  \expandafter\newcommand\csname #1\endcsname[1]{%
    \textcolor{#3}{[\textsf{\footnotesize \textbf{#2:} ##1]}}%
  }
}

\newcommand{\red}[1]{\textcolor{red}{#1}}

\newcommand{\btcb}{\begin{tcolorbox}}
\newcommand{\etcb}{\end{tcolorbox}}
\newcommand{\bbm}{\begin{bmatrix}}
\newcommand{\ebm}{\end{bmatrix}}
\newcommand{\bassume}{\begin{assumption}}
\newcommand{\eassume}{\end{assumption}}
\newcommand{\be}{\begin{equation}}
\newcommand{\ee}{\end{equation}}
\newcommand{\ben}{\begin{equation*}}
\newcommand{\een}{\end{equation*}}
\newcommand{\bea}{\begin{aligned}}
\newcommand{\eea}{\end{aligned}}
\newcommand{\ba}{\begin{equation}\begin{aligned}}
\newcommand{\ea}{\end{aligned}\end{equation}}
\newcommand{\bd}{\begin{definition}}
\newcommand{\ed}{\end{definition}}
\newcommand{\bprop}{\begin{proposition}}
\newcommand{\eprop}{\end{proposition}}
\newcommand{\bt}{\begin{theorem}}
\newcommand{\et}{\end{theorem}}
\newcommand{\bcor}{\begin{corollary}}
\newcommand{\ecor}{\end{corollary}}
\newcommand{\beg}{\begin{example}}
\newcommand{\eeg}{\end{example}}
\newcommand{\bnt}[1]{\begin{namedthm}{#1}}
\newcommand{\ent}{\end{namedthm}}
\newcommand{\blm}{\begin{lemma}}
\newcommand{\elm}{\end{lemma}}
\newcommand{\bp}{\begin{proof}}
\newcommand{\ep}{\end{proof}}
\newcommand{\bpb}{\begin{problem}}
\newcommand{\epb}{\end{problem}}
\newcommand{\benum}{\begin{enumerate}}
\newcommand{\eenum}{\end{enumerate}}
\newcommand{\bitem}{\begin{itemize}}
\newcommand{\eitem}{\end{itemize}}
\definecolor{firebrick}{RGB}{178,34,34}

\def\brst\begin{restatable}
\newcommand{\erst}{\end{restatable}}

\usepackage{threeparttable }
\usepackage{multicol,multirow}
\usepackage[hyphens]{url}

\reviewer{jiayao}{JZ}{NavyBlue}
\reviewer{ting}{TY}{red}
\reviewer{dan}{DR}{blue}

\title{Association between author metadata and acceptance: A feature-rich, matched observational study of a corpus of ICLR submissions between 2017-2022}

\author[1, $\dagger$]{Chang Chen}
\author[2,3, $\dagger$]{Jiayao Zhang}
\author[2]{Dan Roth}
\author[1]{Ting Ye}
\author[4]{Bo Zhang \thanks{Correspondence to Bo Zhang, Vaccine and Infectious Disease Divison, Fred Hutchinson Cancer Center, Seattle, Washington, 98109. Email: bzhang3@fredhutch.org}}

\affil[1]{Department of Biostatistics, University of Washington}
\affil[2]{Department of Computer and Information Science, University of Pennsylvania}
 \affil[3]{Department of Statistics and Data Science, University of Pennsylvania}
\affil[4]{Vaccine and Infectious Disease Division, Fred Hutchinson Cancer Center}
\affil[$\dagger$]{Contributed equally}

\date{}
\begin{document}

\maketitle
\noindent
\textsf{{\bf Abstract}: }%
Many recent studies have probed status bias in the peer-review process of academic journals and conferences. In this article, we investigated the association between author metadata and area chairs' final decisions (Accept/Reject) using our compiled database of $5,313$ borderline submissions to the International Conference on Learning Representations (ICLR) from 2017 to 2022. We carefully defined elements in a cause-and-effect analysis, including the treatment and its timing, pre-treatment variables, potential outcomes and causal null hypothesis of interest, all in the context of study units being textual data and under Neyman and Rubin's potential outcomes (PO) framework. We found some weak evidence that author metadata was associated with articles' final decisions. We also found that, under an additional stability assumption, borderline articles from high-ranking institutions (top-30\% or top-20\%) were less favored by area chairs compared to their matched counterparts. The results were consistent in two different matched designs (odds ratio = $0.82$ [95\% CI: $0.67$ to $1.00$] in a first design and $0.83$ [95\% CI: $0.64$ to $1.07$] in a strengthened design). We discussed how to interpret these results in the context of multiple interactions between a study unit and different agents (reviewers and area chairs) in the peer-review system.

\vspace{0.3 cm}


\section{Introduction}
\label{sec: introduction}

\subsection{Implicit bias in the peer review process}
\label{subsec: implicit bias}
Peer review has been the cornerstone of scientific research. It is important that the peer review process be fair and impartial, especially for early-career researchers. In recent years, the peer review process has been under a lot of scrutiny. For instance, 
in 2014, the organizers of the Conference on Neural Information
Processing Systems (NeurIPS) randomly duplicated
$10\%$ of submissions and assigned them to two independent
sets of reviewers. The study found that $25.9\%$ of these submissions received inconsistent decisions\footnote{See \url{https://inverseprobability.com/2014/12/16/the-nips-experiment}
for more details.}. \citet{nips2014} tracked the fate of submissions rejected in the NeurIPS experiment and found that the peer review process was good at identifying poor papers but fell short of pinpointing good ones. In a similar vein,
\citet{mcgillivray2018uptake} analyzed 128,454 articles in Nature-branded journals and found that authors from less prestigious academic institutions are more likely to choose double-blind review as opposed to single-blind review.
More recently, 
\citet{sun2022does} studied 5,027 papers submitted to the International Conference on Learning Representations (ICLR) and found that scores given to the most prestigious authors significantly decreased after the conference switched its review model from single-blind review to double-blind review. 
\citet{smirnova2022nudging} evaluated a policy that encourages (but did not force) authors to anonymize their submissions and found that the policy increased positive peer reviews by 2.4\% and acceptance by 5.6\% for low-prestige authors while slightly decreased positive peer reviews and acceptance rate for high-prestige authors. Many of these studies identified associations between decision makers' perception of certain aspects of articles' author metadata (e.g., authors' prestige or identity) and final acceptance decisions of these articles, and suggested various forms of implicit bias in the peer review processes, especially among those that adopt a single-blind model.




\subsection{A hypothetical experiment}
\label{subsec: a hypothetical RCT}

In a seminal paper, \cite{bertrand2004emily} measured racial discrimination in labor markets by sending resumes with randomly assigned names, one African American sounding and the other White sounding (e.g., Lakisha versus Emily), to potential employers. \citeauthor{bertrand2004emily}'s \citeyearpar{bertrand2004emily} study was elegant for two reasons. First, it was a randomized experiment that is free of confounding bias, observed or unobserved, {although to what extent the found effect could be attributed to the bias towards applicants' race and ethnicity versus towards other personal traits signaled by the names is unclear}; see, e.g., related discussions in \citet[Section IV]{bertrand2004emily} and \citet{greiner2011causal}. Second, the study illustrated a general strategy to measure the causal effect due to an immutable trait: instead of imagining manipulating the immutable trait itself, the study manipulates employers' \emph{perception} of this immutable trait. 

In a recent high-profile study published in the \emph{Proceedings of the National Academy of Sciences}, \citet{huber2022nobel} designed a field experiment in the similar spirit as \citet{bertrand2004emily}. \citet{huber2022nobel} measured the extent of the \emph{status bias}, defined as a differential treatment of the same paper by prominent versus less established authors in the peer-review process, by randomizing over $3,300$ researchers to one of the three arms: one arm assigned an article with a prestigious author, one arm assigned an anonymized version of the same article, and the other arm assigned the same article but with a less established author. \citet{huber2022nobel} found strong evidence that the prominence of authorship markedly increased the acceptance rate by as much as sixfold, although to what extent this conclusion generalizes to other contexts, e.g., other articles in the same field or articles in other fields, is unclear.

\citeauthor{bertrand2004emily} \citeyearpar{bertrand2004emily} and \citeauthor{huber2022nobel}'s \citeyearpar{huber2022nobel} studies illuminate a randomized experiment that conference organizers and journal editorial offices could carry out, at least hypothetically, in order to understand various forms of bias. For instance, if the policy interest is to evaluate the effect of reviewers' perception of certain aspects of authors (e.g., authors' identity, institution, etc), then a hypothetical experiment would forward to reviewers articles with randomly assigned aspects of interest. Although such an experiment is conceivable, it is difficult to implement due to practical constraints. 

\subsection{A quasi-experimental design using a corpus of ICLR papers}
In the absence of an RCT, a quasi-experimental design aims to fill in the gap by constructing two groups, one treatment group and the other comparison group, that are as similar as possible in pre-treatment variables from retrospective, observational data. Statistical matching is a popular quasi-experimental design device \citep{rosenbaum2002observational,rosenbaum2010design}. In this article, we describe an effort to conduct a matched observational study using state-of-the-art natural language processing tools and study design devices in an effort to investigate the effect of authorship metadata on papers' final decisions.


Our database was constructed from a carefully curated corpus of papers from the International Conference on Learning Representations (ICLR), 
a premium international machine learning conference.
The database is the first of its kind to provide an empirical evidence base for investigating the peer review process. In particular, the database is feature-rich, in the sense that it contains not only explicit/structural features of an article such as its keywords, number of figures and author affiliations, but also more subtle and higher-level features such as topic and textual complexity as reflected by articles' text, and reviewers' sentiment as reflected by their publicly available comments. Building upon discussions of immutable traits regarding human subjects, for instance, those in \citet{greiner2011causal}, we elaborate on the potential outcomes framework \citep{neyman1923application, rubin1974estimating} that facilitates a cause-and-effect analysis between authorship and papers' final decisions; in particular, we will carefully define and state the treatment of interest including its timing, pre-treatment variables, causal identification assumptions, causal null hypothesis to be tested and how to interpret the results, all in the context of study units being textual data.

The conference submission and peer review process consist of multiple steps. For a typical machine learning conference like ICLR, articles need to be submitted by authors before a pre-specified deadline. Valid submissions are then forwarded to a number of reviewers (typically three to four) for feedback and a numerical rating. This part of the peer review process is double-blind so the reviewers and authors in principle do not know each other although in practice, reviewers could identify authors from penmanship or because the authors may upload their articles to the preprint platform \url{arXiv.org}. Authors are then given the chance to answer reviewers' comments and feedback and provide a written rebuttal. Reviewers are allowed to modify their previous ratings taking into account the rebuttal. Finally, an area chair (similar to an associate editor of an academic journal) reviews the article, its ratings and then make a final decision. Submitted articles, author metadata, reviewers' written comments, authors' written rebuttals, reviewers' ratings and area chairs' final decisions are all openly available from the website \url{openreview.net}.

\begin{figure}[t]
	\begin{subfigure}[t]{0.5\columnwidth}
	    \centering
\includegraphics[width=\linewidth]{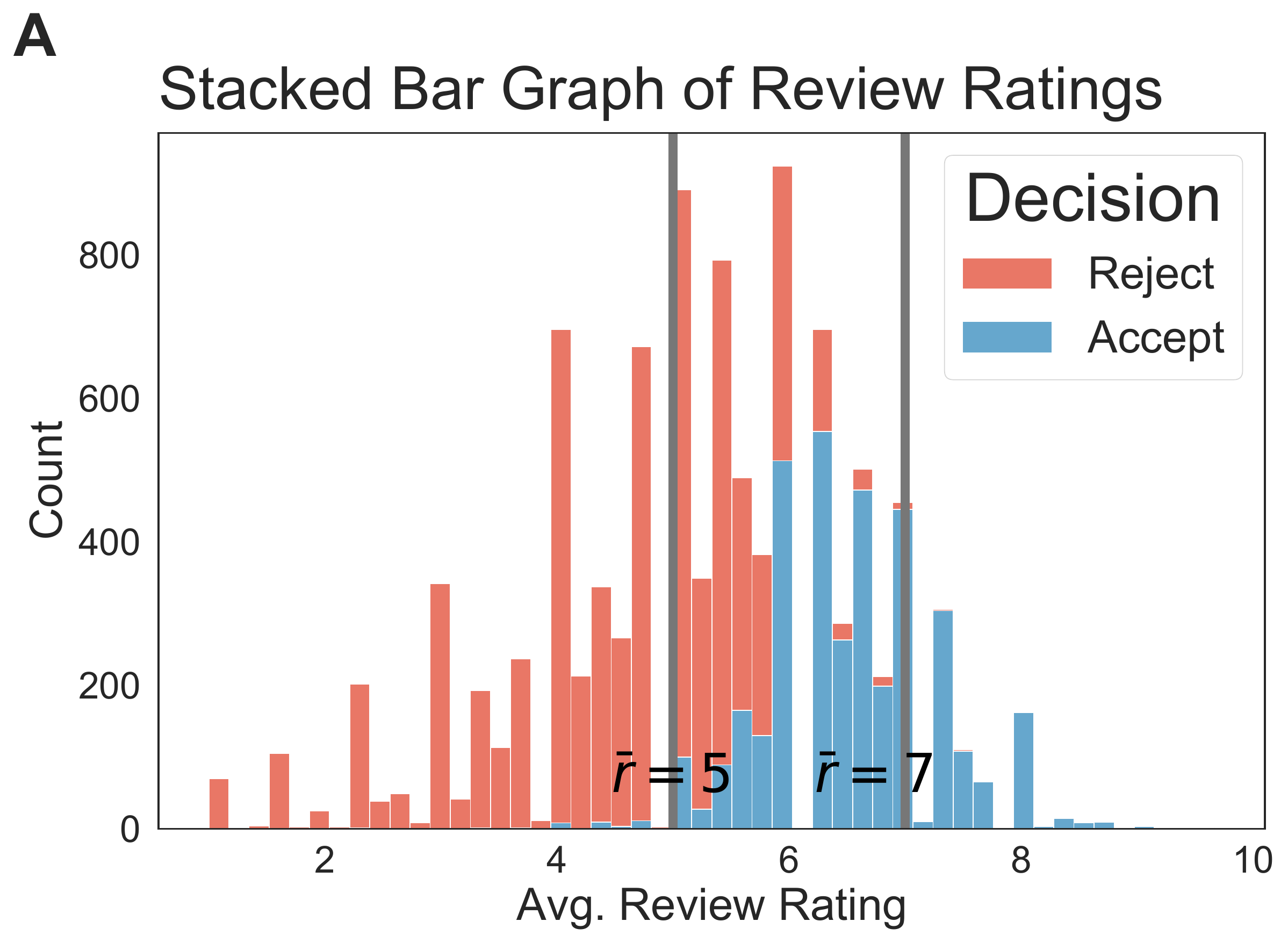}
	\end{subfigure}\hfill
    \begin{subfigure}[t]{0.5\columnwidth}
	    \centering
\includegraphics[width=\linewidth]{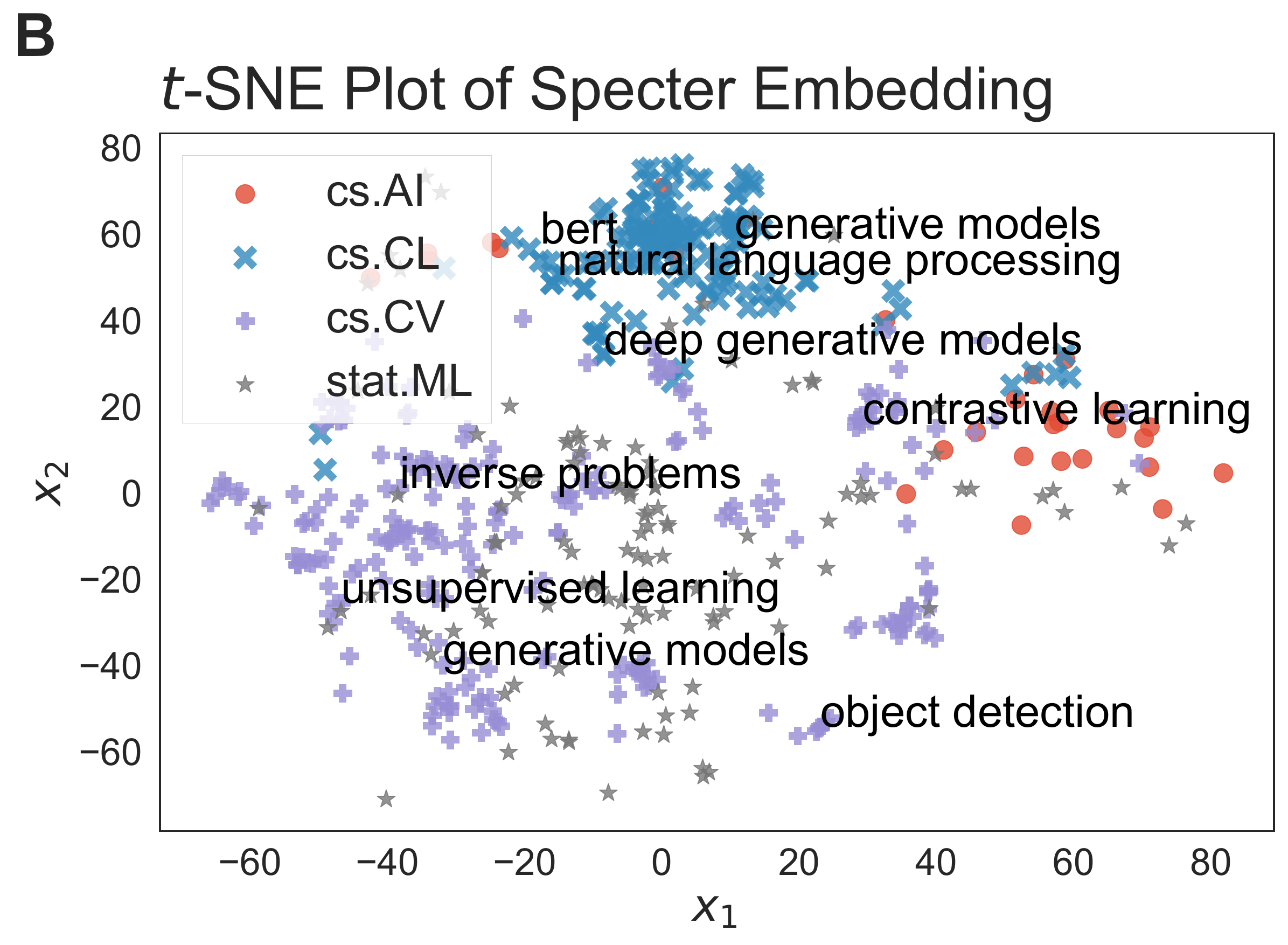}
	\end{subfigure}\\
\begin{subfigure}[t]{0.5\columnwidth}
	    \centering
\includegraphics[width=\linewidth]{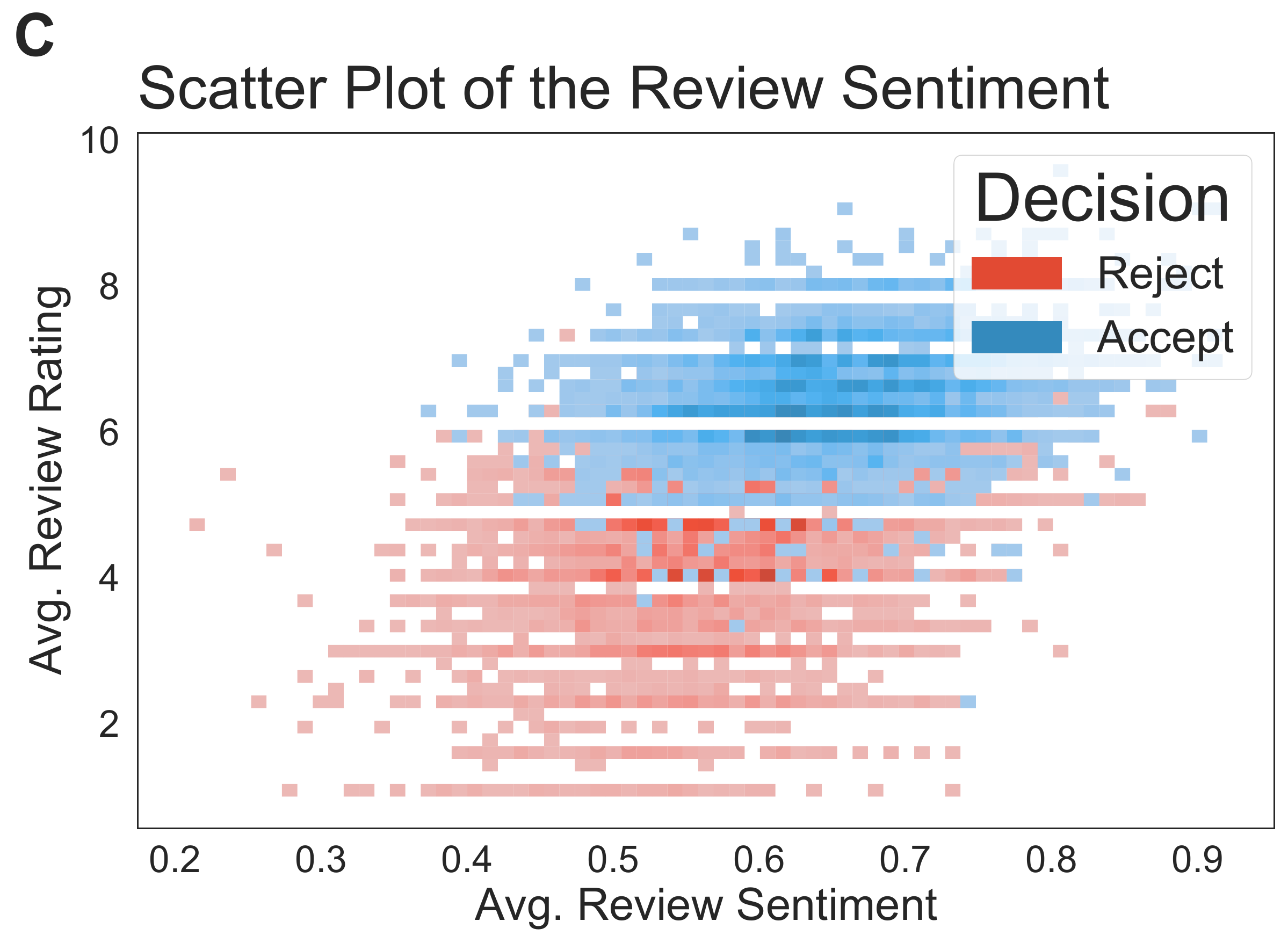}
	\end{subfigure}\hfill
    \begin{subfigure}[t]{0.5\columnwidth}
	    \centering
\includegraphics[width=\linewidth]{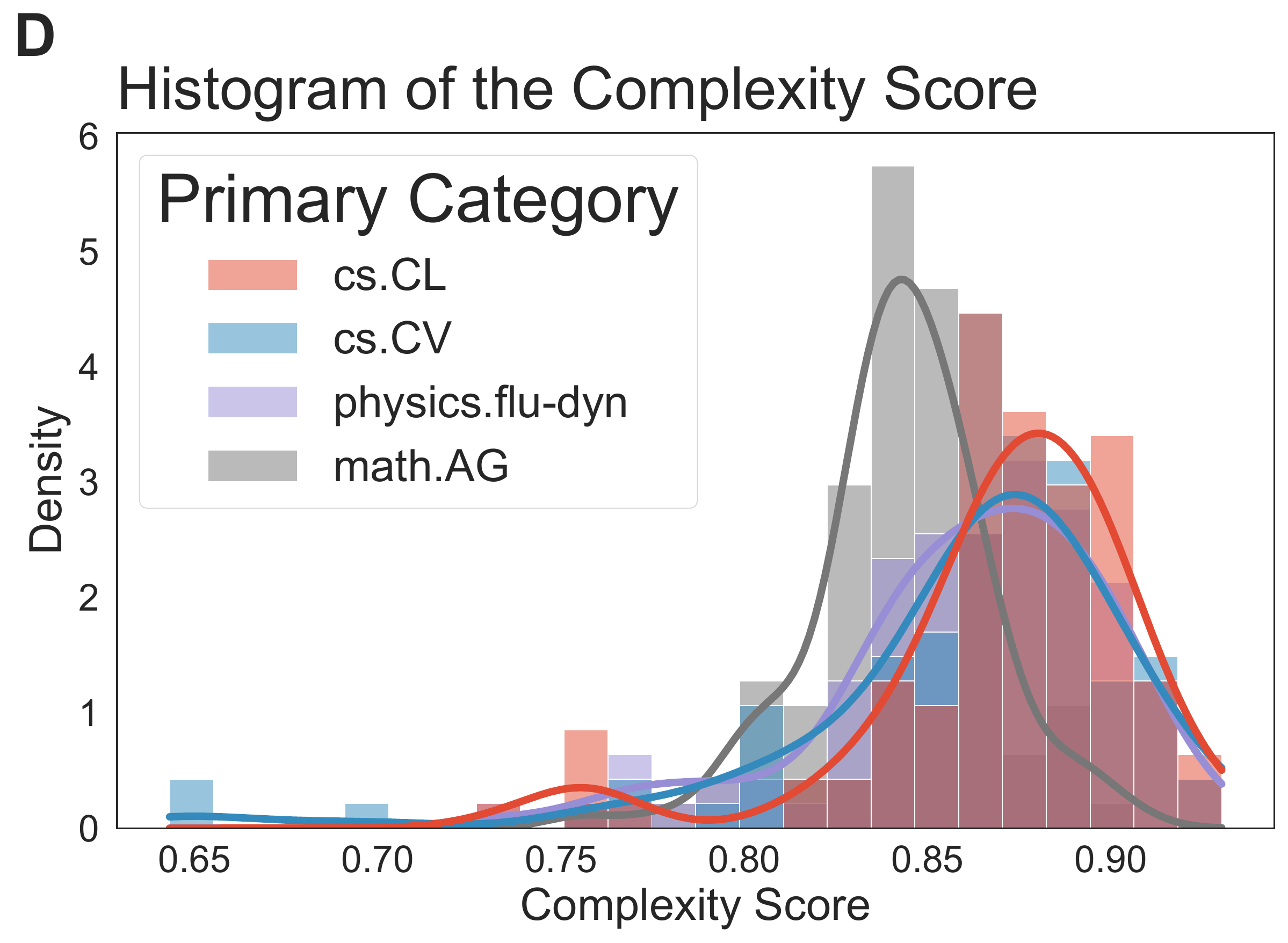}
	\end{subfigure}\\
    \caption{\textbf{Illustration of ratings and high-level textual features.} (A) Stacked bar graphs of acceptance decisions. The decisions of borderline papers (average rating between $5$ and $7$) are not clearcut.
    (B) $T$-SNE plot of
    the Specter embedding together
    with arXiv primary categories and primary keywords among submissions that have been put onto arXiv.
    (C) Sentiment and ratings of reviews. (D) Distribution of complexity scores of a random samples of arXiv articles across four primary categories.}\label{fig:eda}
\end{figure}

Our compiled database allows us to study many different aspects of the peer review process. In this article, we will focus specifically on the last stage of the peer review process and investigate the effect of authorship metadata on area chairs' final decisions. Our focus was motivated by several considerations. First, it is an empirical fact that articles receiving identical or near-identical ratings could receive different final decisions; see, e.g., the stacked bar graph in Panel A of Figure \ref{fig:eda}. It is not unreasonable for authors, especially those who are junior and less established, to wonder if they are fairly treated. Second, any endeavor to establish a cause-and-effect relationship using observational data is challenging because of unmeasured confounding. In our analysis, articles have many implicit features like novelty, thoughtfulness, etc, and these unmeasured confounders could in principle explain away any association between author metadata and final decisions. This problem is greatly attenuated when we focus on area chairs' final decisions and have reviewers' ratings as \emph{pre-treatment} covariates: reviewers' ratings should be a good, albeit not perfect, proxy for the innate quality of an article.

The plan for the rest of the article is as follows. In Section \ref{sec: ICLR data}, we briefly describe our compiled database. In Section \ref{sec: framework}, we lay out the potential outcomes (PO) framework under which we will conduct the matched observational study. Section \ref{sec: study design} describes two concrete study designs, including matching samples and underlying matching algorithms. We also report outcome analysis results in Section \ref{sec: study design}. Section \ref{sec: discussion} discusses how to interpret our findings.


\section{Data: ICLR papers from 2017-2022}
\label{sec: ICLR data}

\subsection{Database construction and structural features}
\label{subsec: data structral features}
We use the ICLR Database\footnote{\url{https://cogcomp.github.io/iclr_database/}} collected and complied
by \citet{ZZDR22}. 
Motivated by the observation that area chairs' final decisions of submissions with an average rating between $5$ to $7$ are not deterministic, as shown in Panel A of Figure~\ref{fig:eda}, we restrict ourselves to this subset of borderline submissions. We first briefly
recall the data collection and cleaning process here for completeness.

The \texttt{OpenReview} API
is
used to crawl data of $10,289$ submissions to 
the International Conference on Learning Representations (ICLR) from $2017$ to $2022$. The crawled data include
(i) submissions and their metadata;
(ii) author metadata, (iii) review/rebuttal data and (iv) area-chair/program chair decision.
Structural features, including the number of sections, figures, tables and bibliographies, are extracted in a relatively straightforward manner. We also extracted and rank self-reported keywords from each submission to form primary and secondary keywords.
Author profiles include an optional self-reported gender; we used the first name dictionary developed by \citet{tsai2016cross}
to provide a numerical score based on the first names of the authors where
$0$ signifies female and $1$ otherwise.
Author profiles are then
augmented via Google Scholar API
to obtain
author citation and $h$-index data.
Author institution is matched using the
domain name of the author email\footnote{
Only domain names are visible from the OpenReview API; other information is masked.}. Although
CSRanking
data is available, it does not
have full coverage of all authors' institutions. As such, we mainly use
the institutional ranking derived from the cumulative number of accepted papers
to the ICLR in the past. For example, the ranking in 2020 of institution A is determined by all papers accepted to ICLR 2017-2020 that have at least one author from it. The review data include rating, confidence and textual reviews. In some years, for example, 2020 and 2022, there are additional assessments such as technical soundness or novelty. Since these additional assessments are not available
for all years, we restrict our attention
to ratings, confidence, and higher-level features derived from textual reviews to be discussed shortly.
Finally, we dichotomize the paper decision by grouping various acceptance designations (spotlight, poster, short talk) into ``Accept'' and ``Reject" or ``invited to workshop track" as ``Reject.''

We also identify if a submission has been posted on the preprint platform \url{arXiv.org}
before the review deadline (i.e., the time reviewers are asked to finalize their reviews)
by (i) searching for five most relevant results based on the title and abstract corresponding to each article from \url{arXiv.org}, (ii) computing the Jacard similarity and normalized Levenshtein similarity between authors, and (iii) calculating the cosine-similarity of the title-abstract embedding. Using the arXiv timestamp, we then identified which submissions were posted prior to the end of the review process.  Among the subset of papers that has arXiv counterparts, we
also obtain their arXiv-related metadata such as primary and secondary categories.



\subsection{Derived higher-level features}

Although the structural features described
so far contain abundant information,
we considered further leveraging language models to extract additional higher-level
features directly from textual data. These higher-level features, like topics, quality of writing and mathematical complexity or rigor, may help quantify residual confounding not captured by structural aspects (e.g., those described in Section \ref{subsec: data structral features}) of an article. Furthermore, in a matched observational study, it is desirable to have a characterization of the ``similarity" among study units to serve as a criterion for matching. Therefore, we derived the following higher-level features and a similarity measure based on embeddings from language models to facilitate our study design.

\subsubsection{SPECTER Embedding}
Our first tool is the SPECTER model
\citep{specter2020cohan}, a BERT-based model \citep{devlin2019bert}
fine-tuned on a scholarly corpus that has
a good track record on summarization tasks (generate abstracts from the main texts) on academic articles. This model takes
the \emph{abstract} of the submissions
and outputs a $784$-dimensional real-valued vector as its representation. Panel B of Figure~\ref{fig:eda}
plots a two-dimensional $t$-SNE \citep{van2008visualizing}
embedding of this representation across
a subset of $535$ submissions that have their arXiv information and primary keyword available. We see that computer vision and computational linguistics articles separate well while general AI and ML articles blend in. In addition,
we sample $9$ primary keywords to overlay on the $t$-SNE embedding. Note that semantically similar
keywords (e.g., the phrases ``BERT'' and ``natural language processing'') generally
have higher proximity under this embedding, which further demonstrate its effectiveness.
We thus use this embedding to (1) perform a $10$-component spectral clustering to assign each submission a
``semantic cluster'' of submissions, and (2) compute the cosine similarity between any two articles.


\paragraph{Sentiment Modeling.}
A RoBERTa model \citep{liu2019roberta} fine-tuned on Twitter sentiments \citep{rosenthal2017semeval}
is used to assign a sentiment
score to textual reviews where $0$ signifies
negative and $1$ positive. We plot
the scatter plot of average sentiment 
and average rating of submissions
with different color signifies different
decisions in Panel C of Figure~\ref{fig:eda}. We observe that the sentiment is highly correlated with the rating while behaves
more volatile when the rating is borderline. This suggests that incorporating
review sentiment may help complement
numerical ratings in the downstream analysis, especially when numerical ratings are borderline and not discriminative.

\paragraph{Complexity Score.}
We use an off-the-shelf fluency model\footnote{\url{https://huggingface.co/prithivida/parrot_paraphraser_on_T5}} derived
from the RoBERTa model \citep{liu2019roberta} to assess sentence-level
fluency and take the average to represent
the complexity of an article, where
$1$ signifies most fluent and easy-to-read while $0$ denotes gibberish-like
sentences.
This fluency score measures how well each article aligns with the English grammar,
and serves as a proxy for an article's heaviness in mathematical notation 
since in-line mathematical notation often disrupts an English sentence's fluency and results in a lower score.
In Panel D of Figure~\ref{fig:eda}, we perform
a sanity-check by randomly sample approximately $100$ arXiv papers from four categories (computational linguistics, computer vision, fluid dynamics, and algebraic geometry) and compute their complexity score. Note that most of the scores
are relatively high, as expected since
academic articles are often relatively well-written. We also observe a discrepancy in that algebraic geometric
papers has its score distribution significantly skewed to the left,
which also aligns with our intuition.
We thus use this complexity score as
a proxy to mathematical complexity and paper readability in our subsequent analysis.


\section{Notation and framework}
\label{sec: framework}
\subsection{Matched cohort study}
\label{subsec: matched obs intro}
In the absence of a well-controlled experiment (e.g., the hypothetical RCT envisioned in Section \ref{subsec: a hypothetical RCT}), observational studies, \emph{up to their important limitations}, provide an alternative to explore a cause-and-effect relationship \citep{cochran1965planning}. In an observational study, study units receiving different levels of treatment may differ systematically in their observed covariates, and this induces the so-called \emph{overt bias} \citep[Section 3]{rosenbaum2002observational}. In our case study, articles with different author metadata, for instance, those with authors from high-ranking versus relatively lower-ranking academic institutions, could differ systematically in topics, keywords, number of figures and equations, among others, and this would invalidate a na\"ive comparison. 

Statistical matching is a commonly used strategy to adjust for confounding in empirical studies \citep{rubin1973matching,rubin1979using,rosenbaum2002observational,rosenbaum2010design,ho2007matching,stuart2010matching}. The ultimate goal of statistical matching is to embed non-randomized, observational data into an approximate randomized controlled experiment by \emph{designing} a matched control (or comparison) group that resembles the treated group in observed pre-treatment covariates by matching on these covariates \citep{rubin1973matching,rubin1979using}, a balancing score derived from these covariates, e.g., \citeauthor{rosenbaum1983central}'s \citeyearpar{rosenbaum1983central} propensity score, or a combination of both \citep{rosenbaum1985constructing}. 

We note that there are multiple ways to adjust for observed covariates and draw causal conclusions under the potential outcomes framework, statistical matching being one of them. Other commonly used methods include weighting, modeling the potential outcomes, and a combination of both. We found a matched observational study particularly suited for our case study for three reasons. First, it facilitates testing Fisher's sharp null hypothesis of no effect, which is an appropriate causal null hypothesis encoding an intriguing notion of \emph{fairness}, as we will discuss in detail in Section \ref{subsec: causal null hypothesis}. Second, a matched design naturally takes into account similarity of textual data (for instance, as measured by cosine similarity based on their embeddings) and is capable of balancing some high-dimensional covariates like keywords in our data analysis. A third strength is mostly stylistic: A matched comparison best resembles \citeauthor{bertrand2004emily}'s \citeyearpar{bertrand2004emily} seminal field experiment and is perhaps the easiest-to-digest way to exhibit statistical analysis results to a non-technical audience.

In the rest of this section, we articulate essential elements in our analysis, including study units, treatment to be hypothetically manipulated, potential outcomes, timing of the treatment, pre-treatment variables and causal null hypothesis of interest. 

\subsection{Study units; treatment and its timing; potential outcomes}
\label{subsec: study units, treatment and PO}
As discussed in detail in \citet{greiner2011causal}, there are two agents in our analysis of the effect of authorship metadata on area chairs' final decisions: an ICLR article peer-reviewed and having received reviewers' ratings, and a decision maker, i.e., an area chair or meta reviewer (AC for short), who assigned the final acceptance status to the article. In our analysis, each study unit is a (peer-reviewed ICLR article, area chair) pair. There are a total of $N = 10,289$ study units in our compiled database, and $5,313$ of them have three or four reviewers and an average rating between $5$ and $7$ and . We will write the $i$-th study unit as $SU_i = (\text{article}_i, \text{AC}_i)$.

We define the treatment of interest as an area chair's perception of a peer-reviewed article's authorship metadata. This definition is modeled after \citet{bertrand2004emily} and \citet{greiner2011causal} and implies the \emph{timing} of the treatment: We imagine a hypothetical randomized experiment where peer-reviewed ICLR articles, whose constituent parts include text, tables, figures, reviewers' ratings and comments, are randomly assigned authorship metadata and presented to the area chair for a final decision. This timing component of the treatment is critical because it implies what are meant to be ``pre-treatment variables" under Neyman-Rubin's causal inference framework, as we will discuss in detail in Section \ref{subsec: observed pre-treatment variables}. 

In principle, the most granular author metadata is a complete list of author names with their corresponding academic or research institutions. Let $\Vec{A} = \Vec{a}$ denote author metadata and $\mathcal{A}$ the set of all possible configurations of author metadata. There is one potential outcome $Y_i(\Vec{a})$ associated with unit $i$ and each $\Vec{a} \in \mathcal{A}$; in words, there is one final decision associated with each peer-reviewed article had the author metadata been $\Vec{a}$. We will assume the consistency assumption so that the observed outcome $Y_i^{\textsf{obs}} = Y_i(\Vec{a}^{\text{obs}})$. One may adopt a variant of the \emph{Stable Unit Treatment Value Assumption} or SUTVA \citep{rubin1980discussion} to reduce the number of potential outcomes. For instance, one may further assume that the potential outcome $Y_i(\Vec{a})$ depends on author metadata $\Vec{a}$ only via authors' academic institutions. Let $f(\cdot)$ denote a mapping from author metadata to authors' academic institutions, then this ``stability" assumption amounts to assuming $Y_i(\Vec{a}) = Y_i(\Vec{a}')$ when $f(\Vec{a}) = f(\Vec{a}')$. We do not \emph{a priori} make such stability assumptions. 

\begin{example}[Field experiment in \citet{bertrand2004emily}]\rm
In \citeauthor{bertrand2004emily}'s \citeyearpar{bertrand2004emily} field experiment, each study unit consists of a resume $i$ and a human resource person reading the resume $i$, i.e., $SU_i = (\text{resume}_i, \text{HR person}_i)$. Treatment is a person's perception of the name on the resume. In this case, $\mathcal{A}$ would consist of all names and $Y_i(A = a)$ is the potential administrative decision had the resume $i$ been associated with name $a$. If we further make the stability assumption that $Y_i(a)$ depends on $a$ only via its race and ethnicity connotation as in \citet{bertrand2004emily} and define $f(a) = 1$ if the name $a$ is African-American sounding and $0$ if it is White sounding, then the set of potential outcomes $Y_i(a),~a \in \mathcal{A}$ would reduce to $\{Y_i(1), Y_i(0)\}$.
\label{ex:Field experiment}
\end{example}

\subsection{Observed and unobserved pre-treatment variables}
\label{subsec: observed pre-treatment variables}
According to \citet{rubin2005causal}, covariates refer to ``variables that take their values before the treatment assignment or, more generally,
simply cannot be affected by the treatment." Below, we briefly review a dichotomy of pre-treatment variables in the context of drawing causal conclusions from textual data \citep{Zhang2022Some}.

In human populations research, pre-treatment variables or covariates are often divided into two broad categories: observed and unobserved; see, e.g., \citet{rosenbaum2002observational,rosenbaum2010design}. A randomized controlled experiment like the one in \citet{bertrand2004emily} had the key advantage of balancing both observed and unobserved confounding, while drawing causal conclusions from observational data inevitably suffers from the concern of unmeasured confounding and researchers often control for a large number of observed covariates in order to alleviate this concern. 

When study units are textual data, \citet{Zhang2022Some} divides observed covariates into two types: \emph{explicit observed covariates} $\bf X^{\textsf{exp}}_{\textsf{obs}}$ that could be derived from textual data at face value, e.g., number of equations, tables and illustrations in the article, and \emph{implicit observed covariates} $\bf X^{\textsf{imp}}_{\textsf{obs}}$ that capture higher-level aspects of textual data, e.g., the topic, flow and novelty of the article. In our case study, we will consider the following explicit observed covariates: year of submission, reviewers' ratings, number of authors, sections, figures and reference, and keywords. We further extracted each article's complexity, topic and reviewers' sentiment using state-of-the-art, natural language processing models as described in Section \ref{sec: ICLR data}.

Unmeasured confounding is a major concern for any attempt to draw a cause-and-effect conclusion from observational data, regardless of the covariance adjustment method. Despite researchers' best intention and effort to control for all relevant pre-treatment variables via matching, there is always a concern about unmeasured confounding bias as we are working with observational data. In our analysis of ICLR papers, we identified two sources of unmeasured confounding. First, there could be residual confounding due to the insufficiency of language models (such as the SPECTER model) in summarizing or extracting implicit observed covariates $\bf X^{\textsf{imp}}_{\textsf{obs}}$ like topics, flow and sentiment. Second, in our analysis, we used numeric ratings from reviewers as a proxy of the quality and novelty of the article. Reviewers' ratings may not be sufficient in summarizing the quality of the articles. Unmeasured confounding may lead to a spurious causal conclusion and researchers routinely examine the robustness of the putative causal conclusion using a sensitivity analysis (see, e.g., \citealp{rosenbaum2002observational,rosenbaum2010design,vanderweele2017sensitivity}, among many others).

\subsection{Causal null hypothesis: A case for Fisher}
\label{subsec: causal null hypothesis}
A causal statement is necessarily a comparison among potential outcomes. In the context of a many-level treatment assignment, Fisher's sharp null hypothesis states the following:
\begin{equation}
    \label{eqn: Fisher's sharp null}
H_{0, \text{sharp}}: \quad  Y_{i}(\Vec{a}) = Y_{i}(\Vec{a}'),\quad \forall i = 1, \dots, N~\text{and}~ \Vec{a},\Vec{a}' \in \mathcal{A}.
\end{equation}
Fisher's sharp null hypothesis prescribes a notion of fairness that, arguably, best suits our vision: area chairs' final decisions of the articles are \emph{irrelevant} of author metadata; in other words, the decision $Y_i(\Vec{a})$ could potentially depend on any substantive aspect of the article $i$, including its topic, quality of writing, reviewers' ratings, etc, but would remain the same had we changed author metadata from $\Vec{a}$ to $\Vec{a}'$.

In addition to Fisher's sharp null hypothesis, Neyman's weak null hypothesis, which states that the sample average treatment effect is zero, is another commonly tested causal null hypothesis. Unlike Fisher's sharp null, Neyman's weak null hypothesis allows perception bias of varying magnitude for all article-AC pairs, as long as these biases would cancel out each other in one way or another. We found this a sub-optimal notion of fairness compared to that encoded by Fisher's sharp null hypothesis, and we will focus on testing Fisher's sharp null hypothesis in our data analysis.

\begin{example}[continues=ex:Field experiment]
\citet{bertrand2004emily} found that White sounding names receive $50$ percent more callbacks for interviews; under the stability assumption discussed in Section \ref{subsec: study units, treatment and PO}, their findings could be interpreted as a causal effect of perceiving White versus African-American sounding names. In the absence of the stability assumption, \citeauthor{bertrand2004emily}'s result could still be interpreted as providing evidence against Fisher's sharp null hypothesis $H_{0, \text{sharp}}$ in its most generic form, although in what specific ways $H_{0, \text{sharp}}$ is violated needs further investigation.
\end{example}

Unlike \citeauthor{bertrand2004emily}'s \citeyearpar{bertrand2004emily} randomized experiment that randomly assigns resume names, our cohort of ICLR articles are not randomly assigned authorship metadata. It is conceivable that articles with more ``prestigious" authors, however one might want to define the concept of ``prestige," could differ systematically in their reviewers' ratings, topics, etc, and this difference in baseline covariates could potentially introduce a spurious association between author metadata and area chairs' final decisions. To overcome this, we embed the observational data into a matched-pair design by constructing $I$ matched pairs, each with two peer-reviewed articles, indexed by $j = 1, 2$, such that these two articles are as similar as possible in their covariates but with different author metadata. Let $\Vec{a}_{ij}$ denote the author metadata associated with article $j$ in the matched pair $i$. Such a matched-pair design enables us to test the following sharp null hypothesis:

\begin{equation}
    \label{eqn: Fisher's sharp null in a matched pair design}
H'_{0, \text{sharp}}: \quad  Y_{ij}(\Vec{a}_{i1}) = Y_{ij}(\Vec{a}_{i2}),\quad \forall i = 1, \dots, I,~j = 1,2.
\end{equation}

We note that $H_{0, \text{sharp}}$ in \eqref{eqn: Fisher's sharp null} implies $H'_{0, \text{sharp}}$ in \eqref{eqn: Fisher's sharp null in a matched pair design}, so rejecting $H'_{0, \text{sharp}}$ would then provide evidence against $H_{0, \text{sharp}}$. Such a design is termed ``near-far" design in the literature \citep{baiocchi2010building,baiocchi2012near} and has been used in a number of empirical studies (see, e.g., \citealp{lorch2012differential,neuman2014anesthesia,mackay2020transesophageal_valve}, among others).

\section{Data analysis: study design and outcome analysis}
\label{sec: study design}

\subsection{A first matched comparison: design \textsf{M1}}
\label{subsec: design M1}
\begin{figure}[H]
    \centering
    \includegraphics[width=\textwidth]{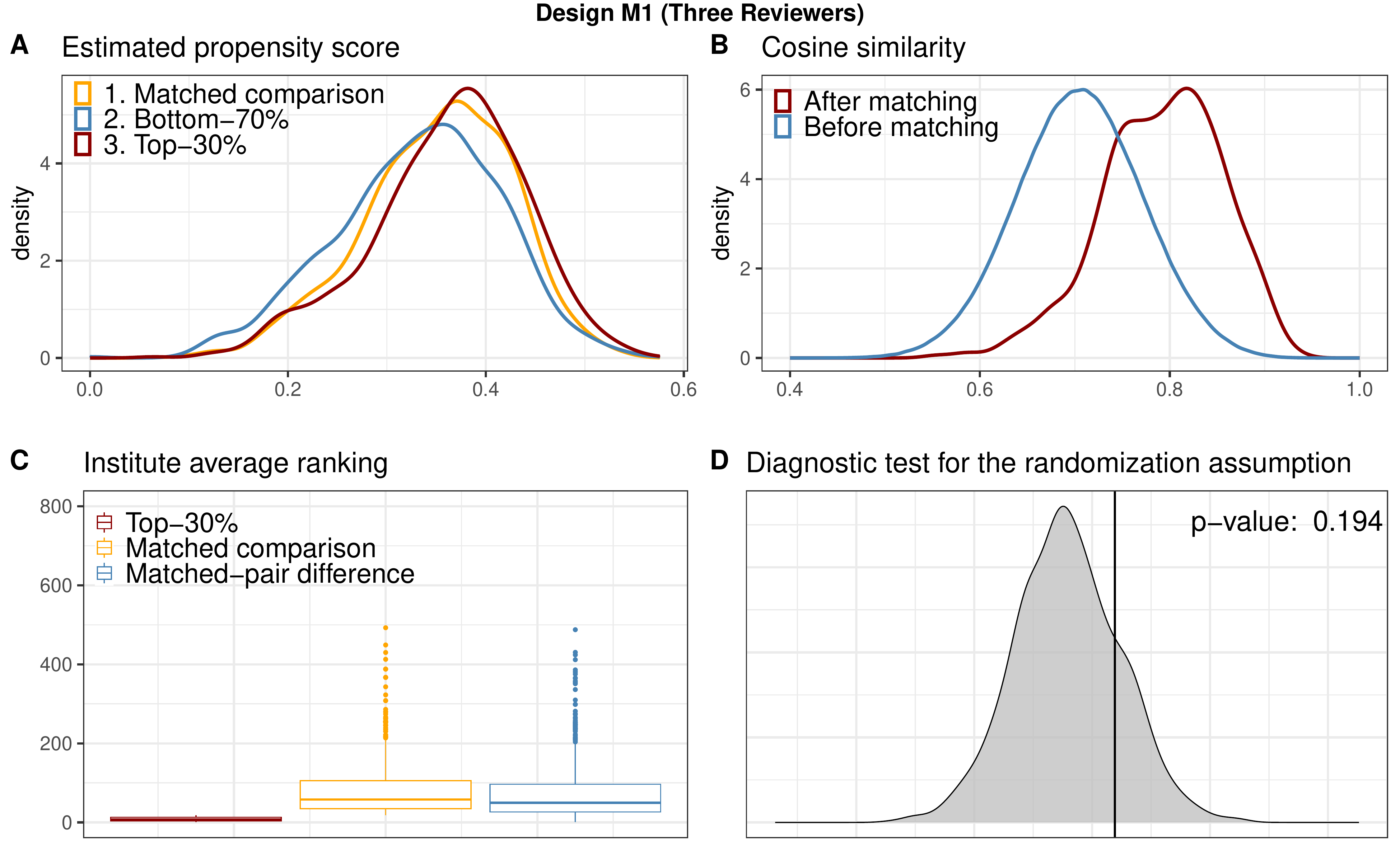}
    \caption{\small \textbf{Diagnostics of design $\textsf{M1}$}. Panel A: Estimated propensity score top-30\% articles, bottom-70\% articles and matched comparison articles. Panel B: Distribution of between-article cosine similarity before (median = 0.71) and after matching (median = 0.80). Panel C: Boxplots of authors' average institution rankings and matched-pair differences in authors' average institution rankings. Panel D: Permutation distribution and the observed test statistic of the classification permutation test. }
    \label{fig: 30 3 reviewers combined}
\end{figure}

We restricted our attention to $5,313$ borderline articles that were peer-reviewed by $3$ or $4$ reviewers and received an average rating between $5$ and $7$. We first considered a study design $\textsf{M1}$ where each matched pair consisted of one article whose authors' average institution ranking was among the top $30\%$ of these $5,313$ submissions and the other article whose authors' average institution ranking was among the bottom $70\%$. Columns 2 and 3 of Table \ref{tbl: balance table main article design M1} summarize the characteristics, including structural features and derived higher-level features, of $1,585$ top-$30\%$ articles and those of the other $3,728$ articles. As one closely examines these two columns, a number of features, including submission year, number of figures, complexity score as judged by the language model, keyword and topic, differ systematically among these submissions. Matching helps remove most of the overt bias: In the matched comparison group (which is a subset of size $n = 1,585$ from the reservoir of $3,728$ articles), standardized mean differences of all but one covariates are less than $0.1$, or one-tenth of one pooled standard deviation. In fact, design $\textsf{M1}$ required near-exact matching on important covariates like reviewers' ratings and year of submission, and achieved near-fine balance for categorical variables like topic cluster and primary keyword \citep{rosenbaum2007minimum}. Algorithms used to construct the matched design $\textsf{M1}$ will be described in detail in Section \ref{subsec: matching algorithm}. Panel C in Figure \ref{fig: 30 3 reviewers combined} assesses how different two articles in each matched pair are in their authors' average institution rankings; see Figure \ref{fig: 30 4 reviewers combined} in Appendix for similar plots in the four-reviewer stratum. The average, within-matched-pair difference in authors' average institution ranking is $74.3$ among $983$ matched pairs in the three-reviewer stratum (median: $49.6$; interquartile range $26.6$-$96.7$) and $99.6$ among $602$ matched pairs in the four-reviewer stratum (median: $63.5$; interquartile range $28.5$-$135.0$). We concluded that there was a sizable difference in author metadata between two articles in each matched pair.

\begin{table}[H]
\centering
\caption{Characteristics of articles before and after matching in the design $\textsf{M1}$.}
\label{tbl: balance table main article design M1}
\resizebox{\textwidth}{!}{
\begin{threeparttable}[h]
\begin{tabular}{lccccccc}
  \hline
 & \multirow{4}{*}{\begin{tabular}{c}Bottom $70\%$\\ Ranking \\Articles \\ (n = 3,728)\end{tabular}} 
 & \multirow{4}{*}{\begin{tabular}{c}Top $30\%$\\ Ranking \\ Articles \\ (n = 1,585)\end{tabular}} 
 & \multirow{4}{*}{\begin{tabular}{c}SMD \\ (Before \\ Matching)\end{tabular}} 
 & \multirow{4}{*}{\begin{tabular}{c}Matched \\Comparison\\ Articles \\ (n = 1,585)\end{tabular}}
 & \multirow{4}{*}{\begin{tabular}{c}SMD \\ (After \\ Matching)\end{tabular}} \\ \\ \\ \\
  \hline
  \multicolumn{6}{c}{\textbf{Conference and Reviewer}} \\
  Year of submission (\%) \\ 
     \hspace{0.5cm}2017 &   109 ( 2.9)  &   129 ( 8.1)  & 0.23 &    92 ( 5.8)  & 0.10 \\ 
     \hspace{0.5cm}2018 &   299 ( 8.0)  &   221 (13.9)  & 0.19 &   201 (12.7)  & 0.04 \\ 
     \hspace{0.5cm}2019 &   520 (13.9)  &   303 (19.1)  & 0.14 &   304 (19.2)  & $<0.01$ \\ 
     \hspace{0.5cm}2020 &   548 (14.7)  &   231 (14.6)  & 0.003 &   234 (14.8)  & $<0.01$ \\ 
     \hspace{0.5cm}2021 &  1200 (32.2)  &   388 (24.5)  & 0.171 &   412 (26.0)  & 0.03 \\ 
     \hspace{0.5cm}2022 &  1052 (28.2)  &   313 (19.7)  & 0.2 &   342 (21.6)  & 0.05 \\ 
  Reviewer Ratings \\
  \hspace{0.5cm}Reviewer I &  6.99 (0.92) &  6.99 (0.91) &  0.008 &  6.97 (0.90) & 0.02 \\ 
  \hspace{0.5cm}Reviewer II &  6.12 (0.80) &  6.12 (0.79) &  0.001 &  6.12 (0.78) & $<0.01$ \\ 
  \hspace{0.5cm}Reviewer III &  5.21 (1.08) &  5.18 (1.09) &  0.031 &  5.19 (1.10) & 0.01 \\
  \hspace{0.5cm}Reviewer IV* & 4.71 (1.06) &  4.74 (1.11) &  0.031 &  4.74 (1.09) & $<0.01$ \\ 
  Reviewer Sentiment \\
 \hspace{0.5cm}Reviewer I &  0.75 (0.10) &  0.75 (0.11) &  0.026 &  0.75 (0.11) & 0.03 \\ 
  \hspace{0.5cm}Reviewer II &  0.64 (0.09) &  0.64 (0.09) &  0.021 &  0.64 (0.09) & 0.04 \\ 
  \hspace{0.5cm}Reviewer III &  0.55 (0.10) &  0.54 (0.10) &  0.081 &  0.54 (0.10) & 0.01 \\
  \hspace{0.5cm}Reviewer IV* &  0.49 (0.09) &  0.50 (0.09) &  0.062 &  0.50 (0.09) & 0.01\\
  \multicolumn{6}{c}{\textbf{Article Metadata}} \\
  No. Author &  4.21 (1.67) &  4.17 (1.69) &  0.024 &  4.13 (1.67) & 0.03 \\ 
  No. Figure & 13.71 (7.26) & 12.55 (7.55) &  0.156 & 12.42 (6.60) & 0.02 \\ 
  No. Reference & 42.53 (16.59) & 42.19 (16.93) &  0.020 & 40.98 (14.94) & 0.07 \\ 
  No. Section & 19.94 (7.16) & 19.96 (7.11) &  0.003 & 19.74 (6.90) & 0.03 \\ 
  \multicolumn{6}{c}{\textbf{Complexity, Topics and Keywords}} \\
   Complexity &  0.84 (0.03) &  0.85 (0.03) &  0.285 &  0.85 (0.03) & 0.06 \\ 
  Topic cluster$^\dagger$ (\%)  \\ 
     \hspace{0.5cm} {RL/Meta Learning/Robustness}&   367 ( 9.8)  &   113 ( 7.1)  & 0.097 &   113 ( 7.1)  & 0 \\ 
     \hspace{0.5cm} {RL/CV/Robustness}&   298 ( 8.0)  &    80 ( 5.0)  & 0.122 &    80 ( 5.0)  & 0 \\ 
     \hspace{0.5cm} {DL/GM/CNN}&   345 ( 9.3)  &   147 ( 9.3)  & 0 &   147 ( 9.3)  & 0 \\ 
     \hspace{0.5cm} {DL/RNN/GNN}&   365 ( 9.8)  &   133 ( 8.4)  & 0.049 &   133 ( 8.4)  & 0 \\ 
     \hspace{0.5cm} {DL/Optimization/Generalization}&   399 (10.7)  &   126 ( 7.9)  & 0.097 &   126 ( 7.9)  & 0 \\ 
     \hspace{0.5cm} {DL/Robustness/Adversarial Examples}&   445 (11.9)  &   270 (17.0)  & 0.145 &   270 (17.0)  & 0 \\ 
     \hspace{0.5cm} {DL/RL/Unsupervised Learning/GM}&   319 ( 8.6)  &   143 ( 9.0)  & 0.014 &   143 ( 9.0)  & 0 \\ 
     \hspace{0.5cm} {DL/Multi-Agent or Model-Based RL/IL}&   475 (12.7)  &   209 (13.2)  & 0.015 &   209 (13.2)  & 0 \\ 
     \hspace{0.5cm} {DL/Federated or Distributed Learning}&   370 ( 9.9)  &   260 (16.4)  & 0.193 &   260 (16.4)  & 0 \\ 
     \hspace{0.5cm} {GM/GAN/VAE}&   345 ( 9.3)  &   104 ( 6.6)  & 0.1 &   104 ( 6.6)  & 0 \\
  Primary keyword (\%) &   \\ 
      \hspace{0.5cm}NA &  950 (25.5)  &   347 (21.9)  & 0.085 &   368 (23.2)  & 0.03 \\ 
      \hspace{0.5cm}Other &   794 (21.3)  &   292 (18.4)  & 0.073 &   312 (19.7)  & 0.03 \\ 
      \hspace{0.5cm}Deep learning &   393 (10.5)  &   242 (15.3)  & 0.144 &   238 (15.0)  & 0.01 \\ 
      \hspace{0.5cm}Reinforcement learning &   290 ( 7.8)  &   183 (11.5)  & 0.126 &   181 (11.4)  & $<0.01$ \\ 
      \hspace{0.5cm}Graph neural networks &   145 ( 3.9)  &    41 ( 2.6)  & 0.073 &    39 ( 2.5)  & $<0.01$ \\ 
      \hspace{0.5cm}Representation learning &   109 ( 2.9)  &    40 ( 2.5)  & 0.025 &    39 ( 2.5)  & $<0.01$ \\ 
      \hspace{0.5cm}Generative models &    89 ( 2.4)  &    35 ( 2.2)  & 0.013 &    38 ( 2.4)  & 0.01 \\ 
      \hspace{0.5cm}Meta-learning &    79 ( 2.1)  &    34 ( 2.1)  & 0 &    33 ( 2.1)  & $<0.01$ \\ 
      \hspace{0.5cm}Self-supervised learning &    72 ( 1.9)  &    23 ( 1.5)  & 0.031 &    24 ( 1.5)  & $<0.01$ \\ 
      \hspace{0.5cm}Unsupervised learning &    70 ( 1.9)  &    43 ( 2.7)  & 0.053 &    32 ( 2.0)  & 0.05 \\ 
      \hspace{0.5cm}Neural networks &    62 ( 1.7)  &    30 ( 1.9)  & 0.015 &    25 ( 1.6)  & 0.02 \\ 
      \hspace{0.5cm}Generative adversarial networks &    56 ( 1.5)  &    14 ( 0.9)  & 0.055 &    18 ( 1.1)  & 0.02 \\ 
      \hspace{0.5cm}Optimization &    43 ( 1.2)  &    26 ( 1.6)  & 0.034 &    26 ( 1.6)  & 0 \\ 
      \hspace{0.5cm}Variational inference &    39 ( 1.0)  &     8 ( 0.5)  & 0.058 &    11 ( 0.7)  & 0.02 \\
      \hspace{0.5cm}Transformer &    37 ( 1.0)  &    20 ( 1.3)  & 0.028 &    15 ( 0.9)  & 0.04 \\ 
      \hspace{0.5cm}Generalization &    36 ( 1.0)  &    32 ( 2.0)  & 0.082 &    22 ( 1.4)  & 0.05 \\ 
     \multicolumn{6}{c}{\textbf{Decision}} \\
       Acceptance (\%) &    1928 ( 51.7) &  811 ( 51.2) & & 851 ( 53.7)\\
     
   \hline 
\end{tabular}
\begin{tablenotes}
       \item [$^\ast$] \textsf{Reviewer IV's rating and sentiment results are derived from articles within the stratum of four reviewers. 
       \item [$^\dagger$] RL: Reinforcement Learning; GM: Generative Models; CV: Computer Vision; CNN: Convolutional Neural Nets; RNN: Recurrent Neural Nets; GNN: Graph Neural Nets; IL: Imitation Learning; GAN: Generative Adversarial Nets; VAE: Variational Auto-Encoder. Note that the description is not exhaustive.}
     \end{tablenotes}
\end{threeparttable}}
\end{table}


To further demonstrate two groups are well comparable, Panel A of Figure~\ref{fig: 30 3 reviewers combined} displays the distribution of the estimated ``propensity score," defined as the probability that authors' average ranking of an article was among top $30\%$, in each of the following three groups: top-30\% articles (red), bottom-70\% articles (blue), and matched comparison articles (yellow), all in the three-reviewer stratum. Similar plots for articles with four reviewers can be found in Appendix. As is evident from the figure, the propensity score distribution of the matched comparison articles is more similar to that of the top-30\% articles. Panel B of Figure \ref{fig: 30 3 reviewers combined} further plots the cosine similarity calculated from the raw textual data of each article. It is also evident that two articles in the same matched pair now have improved cosine similarity compared to that from two randomly drawn articles prior to matching. Our designed matched comparison $\textsf{M1}$ appears to be well balanced in many observed covariates and resembles a hypothetical RCT where we randomly assign author metadata. 

The question remains as to whether the balance is sufficiently good compared to an authentic RCT and could justify randomization inference. We conducted a formal diagnostic test using \citeauthor{gagnon2019classification}'s \citeyearpar{gagnon2019classification} classification permutation test (CPT) based on random forests to test the randomization assumption for the matched cohort. The randomization assumption cannot be rejected in either the three-reviewer or four-reviewer stratum (p-value = $0.194$ and $0.641$, respectively); see the null distribution and test statistic in Panel D of Figure \ref{fig: 30 3 reviewers combined} and Figure \ref{fig: 30 4 reviewers combined} in Appendix. 

\begin{table}[ht]
\centering
\caption{Contingency table, p-values (exact, two-sided McNemar's test) and odds ratio \citep{Fay_exact2by2} of $1,585$ matched pairs in the design $\textsf{M1}$ and $1051$ matched pairs in a strengthened design $\textsf{M2}$. Defining and interpreting the odds ratio and its confidence interval requires an additional ``stability" assumption discussed in Section \ref{subsec: study units, treatment and PO}.}
\label{tbl: mcnemar table}
\begin{tabular}{cccccc}
  \hline
  \multirow{2}{*}{Panel A: Design $\textsf{M1}$} & &\multicolumn{2}{c}{Comparison Articles} & &Odds Ratio\\
 & &Accepted & Rejected &P-value & 95\% CI\\ 
  \hline
\multirow{2}{*}{Top-30\% Articles} &Accepted & 633 & 178 & \multirow{2}{*}{0.050} & 0.82 \\ 
  & Rejected & 218 & 556 & & [0.67, 1.00]\\ 
   \hline

 \multirow{2}{*}{Panel B: Design $\textsf{M2}$} & &\multicolumn{2}{c}{Comparison Articles}\\
 & &Accepted & Rejected \\ 
  \hline
\multirow{2}{*}{Top-30\% Articles} &Accepted & 443 & 115 & \multirow{2}{*}{0.149} & 0.83 \\
  & Rejected & 139 & 354 & &[0.64, 1.07]\\ 
  \hline
   
\end{tabular}
\end{table}

Panel A of Table \ref{tbl: mcnemar table} summarizes the outcomes of $1,585$ matched pairs of two articles. We tested Fisher's sharp null hypothesis $H'_{0, \text{sharp}}$ reviewed and discussed in Section \ref{subsec: causal null hypothesis} using a two-sided, exact McNemar's test \citep{Fay_exact2by2} and obtained a p-value of $0.050$, which suggested some weak evidence that authorship metadata was associated with AC's final decisions. Under an additional stability assumption stating that the potential acceptance status $Y(\Vec{a})$ depended on author metadata $\Vec{a}$ only via authors' average institution ranking and remained unchanged when the average ranking is among the top-30\% or among the bottom-70\%, we estimated the odds ratio to be $0.82$ (95\% CI: [0.67, 1.00]), providing some weak evidence that borderline articles from top-30\% institutions were \emph{less} favored by area chairs compared to their counterparts from the comparison group. Note that a na\"ive, unadjusted comparison between the top-30\% and bottom-70\% borderline articles would in fact mask any difference (acceptance rate: 51.2\% versus 51.7\% before matching).

Our analysis seems to defy the common wisdom that there is a ``status bias" favoring higher-profile authors. It is then natural to ask, if author metadata were even more drastically different, shall we then see some evidence for status bias that would better align with previous findings? This inquiry led to a second, strengthened design $\textsf{M2}$.

\subsection{A strengthened design \textsf{M2}}
\label{subsec: design M2}
\citet{baiocchi2010building} first considered ``strengthening" an encouragement variable (differential distance) in the context of an instrumental variable analysis of the effect of low-level versus high-level neonatal intensive care units (NICUs) on mortality rate. In their analysis, \citet{baiocchi2010building} compared mothers who lived near to and far from a high-level NICU and ``strengthened" the comparison by restricting their analysis to a smaller subset of comparable mothers who lived very near to and very far from a high-level NICU. We adopted their idea here and constructed a strengthened design $\textsf{M2}$ where one article in each matched pair is a top-20\% article (as opposed to a top-30\% article in design $\textsf{M1}$) and the other matched comparison article was from the reservoir of bottom-80\% articles. We also added a ``dose-caliper" in the statistical matching algorithm to further separate the average ranking within each matched pair; see Section \ref{subsec: matching algorithm} for details.

\begin{figure}[H]
    \centering
    \includegraphics[width=\textwidth]{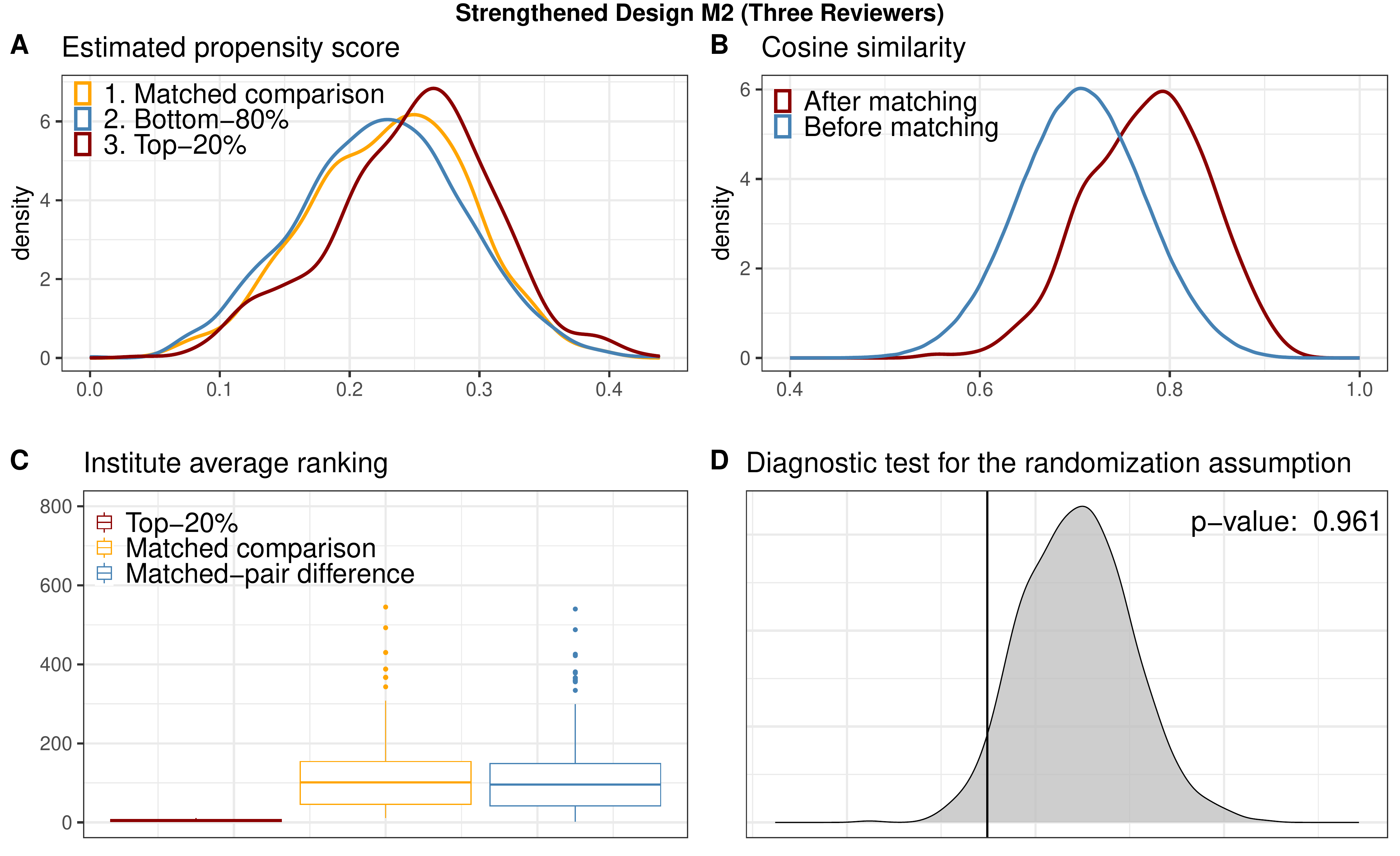}
   \caption{\small \textbf{Diagnostics of design $\textsf{M2}$}. Panel A: Estimated propensity score top-30\% articles, bottom-70\% articles and matched comparison articles. Panel B: Distribution of between-article cosine similarity before (median = 0.71) and after matching (median = 0.78). Panel C: Boxplots of authors' average institution rankings and matched-pair differences in authors' average institution rankings. Panel D: Permutation distribution and the observed test statistic of the classification permutation test. }
    \label{fig: 20 3 reviewers combined}
\end{figure}

A total of $1,051$ matched pairs were formed. Panel C of Figure \ref{fig: 20 3 reviewers combined} summarizes the within-matched-pair difference in authors' average institution ranking across $1,051$ matched pairs in the strengthened design $\textsf{M2}$. In this strengthened design, the average difference now increases from $74.3$ (as in the design \textsf{M1}) to $108.8$. Importantly, the cohort of top-20\% articles and their matched comparison group are still comparable in all baseline covariates, as summarized in Table \ref{tbl: balance table design M2}. Similar to the design \textsf{M1}, we cannot reject the randomization assumption based on the classification permutation test (p-value = $0.961$).

Panel B of Table \ref{tbl: mcnemar table} summarizes the outcomes of $1,051$ matched pairs of two articles in the strengthened design $\textsf{M2}$. Under the additional stability assumption, we obtained a near identical point estimate for odds ratio (OR = $0.83$ in $\textsf{M2}$ versus $0.82$ in $\textsf{M1}$), though the estimate is slightly less precise as a result of a smaller sample size ($1,051$ in \textsf{M2} versus $1,585$ in \textsf{M1}). Consistent results across two study designs help reinforce the conclusion that we did not find evidence supporting a ``status bias" favoring authors from high-ranking institutions in this cohort of borderline articles.

\begin{table}[H]
\centering
\caption{Characteristics of articles before and after matching in a strengthened design $\textsf{M2}$.}
\label{tbl: balance table design M2}
\resizebox{\textwidth}{!}{
\begin{threeparttable}[h]
\label{tbl: balance table supplement}
\begin{tabular}{lccccccc}
  \hline
 & \multirow{3}{*}{\begin{tabular}{c}All Control\\ Articles \\ (n = 4,262)\end{tabular}} 
 & \multirow{3}{*}{\begin{tabular}{c}All Treated\\ Articles \\ (n = 1,051)\end{tabular}} 
 & \multirow{3}{*}{\begin{tabular}{c}SMD \\ (Before \\ Matching)\end{tabular}} 
 & \multirow{3}{*}{\begin{tabular}{c}Matched Control\\ Articles \\ (n = 1,051)\end{tabular}}
 & \multirow{3}{*}{\begin{tabular}{c}SMD \\ (After \\ Matching)\end{tabular}} \\ \\ \\ 
  \hline
  \multicolumn{6}{c}{\textbf{Conference and Reviewer}} \\
  Year of submission (\%) \\ 
     \hspace{0.5cm}2017 &   144 ( 3.4)  &    94 ( 8.9)  & 0.23 &    77 ( 7.3)  & 0.07 \\ 
     \hspace{0.5cm}2018 &   380 ( 8.9)  &   140 (13.3)  & 0.14 &   126 (12.0)  & 0.04 \\ 
     \hspace{0.5cm}2019 &   614 (14.4)  &   209 (19.9)  & 0.146 &   201 (19.1)  & 0.02 \\ 
     \hspace{0.5cm}2020 &   621 (14.6)  &   158 (15.0)  & 0.011 &   157 (14.9)  & $<0.01$ \\ 
     \hspace{0.5cm}2021 &  1343 (31.5)  &   245 (23.3)  & 0.185 &   272 (25.9)  & 0.06 \\ 
     \hspace{0.5cm}2022 &  1160 (27.2)  &   205 (19.5)  & 0.183 &   218 (20.7)  & 0.03 \\ 
  Reviewer Ratings \\
  \hspace{0.5cm}Reviewer I &  6.98 (0.92) &  7.04 (0.91) &  0.059 &  7.01 (0.89) & 0.02 \\ 
  \hspace{0.5cm}Reviewer II &  6.12 (0.79) &  6.13 (0.79) &  0.017 &  6.11 (0.78) & 0.03 \\ 
  \hspace{0.5cm}Reviewer III &  5.21 (1.07) &  5.16 (1.11) &  0.042 &  5.17 (1.09) & $<0.01$ \\
  \hspace{0.5cm}Reviewer IV* &  4.71 (1.06) &  4.77 (1.12) &  0.052 &  4.74 (1.13) & 0.02\\ 
  Reviewer Sentiment \\
 \hspace{0.5cm}Reviewer I &  0.75 (0.10) &  0.76 (0.11) &  0.051 &  0.75 (0.10) & 0.10 \\ 
  \hspace{0.5cm}Reviewer II &  0.64 (0.09) &  0.64 (0.09) &  0.005 &  0.64 (0.09) & 0.08 \\ 
  \hspace{0.5cm}Reviewer III &  0.55 (0.10) &  0.54 (0.10) &  0.062 &  0.54 (0.10) & 0.03 \\
  \hspace{0.5cm}Reviewer IV* &  0.49 (0.09) &  0.50 (0.09) &  0.120 &   393 & 0.06\\
  \multicolumn{6}{c}{\textbf{Article Metadata}} \\
  No. uthor &  4.21 (1.67) &  4.17 (1.72) &  0.023 &  4.19 (1.66) & 0.01 \\ 
  No. Figure & 13.60 (7.41) & 12.41 (7.10) &  0.164 & 12.32 (6.45) & 0.01 \\ 
  No. Reference & 42.56 (16.61) & 41.91 (17.02) &  0.039 & 40.50 (15.12) & 0.08 \\ 
  No. Section & 19.94 (7.16) & 19.96 (7.06) &  0.003 & 19.54 (6.94) & 0.06 \\
  
  \multicolumn{6}{c}{\textbf{Complexity, Topics and Keywords}} \\
   Complexity &  0.84 (0.03) &  0.85 (0.03) &  0.308 &  0.85 (0.03) & 0.15 \\ 
  Topic cluster $^\dagger$ (\%)  \\ 
     \hspace{0.5cm} {RL/Meta Learning/Robustness}&   404 ( 9.5)  &    76 ( 7.2)  & 0.083 &    76 ( 7.2)  & 0 \\ 
     \hspace{0.5cm} {RL/CV/Robustness}&   328 ( 7.7)  &    50 ( 4.8)  & 0.12 &    50 ( 4.8)  & 0 \\ 
     \hspace{0.5cm} {DL/GM/CNN}&   393 ( 9.2)  &    99 ( 9.4)  & 0.007 &    99 ( 9.4)  & 0 \\ 
     \hspace{0.5cm} {DL/RNN/GNN}&   408 ( 9.6)  &    90 ( 8.6)  & 0.035 &    90 ( 8.6)  & 0 \\ 
     \hspace{0.5cm} {DL/Optimization/Generalization}&   439 (10.3)  &    86 ( 8.2)  & 0.073 &    86 ( 8.2)  & 0 \\ 
     \hspace{0.5cm} {DL/Robustness/Adversarial Examples}&   551 (12.9)  &   164 (15.6)  & 0.077 &   164 (15.6)  & 0 \\ 
     \hspace{0.5cm} {DL/RL/Unsupervised Learning/GM}&   365 ( 8.6)  &    97 ( 9.2)  & 0.021 &    97 ( 9.2)  & 0 \\ 
     \hspace{0.5cm} {DL/Multi-Agent or Model-Based RL/IL}&   545 (12.8)  &   139 (13.2)  & 0.012 &   139 (13.2)  & 0 \\ 
     \hspace{0.5cm} {DL/Federated or Distributed Learning}&   435 (10.2)  &   195 (18.6)  & 0.241 &   195 (18.6)  & 0 \\ 
     \hspace{0.5cm} {GM/GAN/VAE}&   394 ( 9.2)  &    55 ( 5.2)  & 0.155 &    55 ( 5.2)  & 0 \\
  Primary keyword (\%) &   \\ 
      \hspace{0.5cm}NA &  1084 (25.4)  &   213 (20.3)  & 0.122 &   220 (20.9)  & 0.01 \\ 
      \hspace{0.5cm}Other &   891 (20.9)  &   195 (18.6)  & 0.058 &   195 (18.6)  & 0 \\ 
      \hspace{0.5cm}Deep learning &   460 (10.8)  &   175 (16.7)  & 0.172 &   178 (16.9)  & $<0.01$ \\ 
      \hspace{0.5cm}Reinforcement learning &   353 ( 8.3)  &   120 (11.4)  & 0.104 &   122 (11.6)  & $<0.01$ \\ 
      \hspace{0.5cm}Graph neural networks &   162 ( 3.8)  &    24 ( 2.3)  & 0.087 &    24 ( 2.3)  & 0 \\ 
      \hspace{0.5cm}Representation learning &   127 ( 3.0)  &    22 ( 2.1)  & 0.057 &    22 ( 2.1)  & 0 \\ 
      \hspace{0.5cm}Generative models &    99 ( 2.3)  &    25 ( 2.4)  & 0.007 &    26 ( 2.5)  & $<0.01$ \\ 
      \hspace{0.5cm}Meta-learning &    89 ( 2.1)  &    24 ( 2.3)  & 0.014 &    23 ( 2.2)  & $<0.01$ \\ 
      \hspace{0.5cm}Self-supervised learning &    80 ( 1.9)  &    15 ( 1.4)  & 0.039 &    14 ( 1.3)  & $<0.01$ \\ 
      \hspace{0.5cm}Unsupervised learning &    79 ( 1.9)  &    34 ( 3.2)  & 0.083 &    26 ( 2.5)  & 0.04 \\ 
      \hspace{0.5cm}Neural networks &    73 ( 1.7)  &    19 ( 1.8)  & 0.008 &    17 ( 1.6)  & 0.02 \\ 
      \hspace{0.5cm}Generative adversarial networks &    62 ( 1.5)  &     8 ( 0.8)  & 0.066 &    10 ( 1.0)  & 0.02 \\ 
      \hspace{0.5cm}Generalization &    48 ( 1.1)  &    20 ( 1.9)  & 0.066 &    16 ( 1.5)  & 0.03 \\ 
      \hspace{0.5cm}Optimization &    47 ( 1.1)  &    22 ( 2.1)  & 0.08 &    22 ( 2.1)  & 0 \\ 
      \hspace{0.5cm}Variational inference &    44 ( 1.0)  &     3 ( 0.3)  & 0.087 &     7 ( 0.7)  & 0.05 \\ 
      
       \multicolumn{6}{c}{\textbf{Decision}} \\
       Acceptance (\%) &    2181 ( 51.2) &  558 ( 53.1) & & 582 ( 55.4)\\
     
   \hline
\end{tabular}
\begin{tablenotes}
       \item [$^\ast$] \textsf{Reviewer IV's rating and sentiment results are derived from articles within the stratum of four reviewers. 
       \item [$^\dagger$] RL: Reinforcement Learning; GM: Generative Models; CV: Computer Vision; CNN: Convolutional Neural Nets; RNN: Recurrent Neural Nets; GNN: Graph Neural Nets; IL: Imitation Learning; GAN: Generative Adversarial Nets; VAE: Variational Auto-Encoder. Note that the description is not exhaustive.}
     \end{tablenotes}
\end{threeparttable}}
\end{table}

\subsection{Matching algorithm: matching one sample according to multiple criteria}
\label{subsec: matching algorithm}
The matched sample $\textsf{M1}$ displayed in Table \ref{tbl: balance table main article design M1} was constructed using an efficient, network-flow-based optimization algorithm built upon a tripartite network \citep{zhang2021matching} as opposed to a traditional, bipartite network \citep{rosenbaum1989optimal}. Compared to a bipartite network, a tripartite network structure helps separate two tasks in the design of an observational study: (1) constructing closely matched pairs and (2) constructing a well balanced matched sample. A detailed account of the algorithm can be found in \citet{zhang2021matching}; below, we described how we designed the cost in the tripartite network and achieved the features of $\textsf{M1}$ and $\textsf{M2}$ described in Section \ref{subsec: design M1} and \ref{subsec: design M2}.

\begin{figure}[ht]
\centering
\begin{tikzpicture}[thick, color = black,
  fsnode/.style={circle, fill=black, inner sep = 0pt, minimum size = 5pt},
  ssnode/.style={circle, fill=black, inner sep = 0pt, minimum size = 5pt},
  shorten >= 3pt,shorten <= 3pt
]


\begin{scope}[start chain=going below,node distance=7mm]
\foreach \i in {1,2,3}
  \node[fsnode,on chain] (r\i) [label=above left: {\small$\gamma_\i$} ] {};
\end{scope}

\begin{scope}[xshift=3cm,yshift=0cm,start chain=going below,node distance=7mm]
\foreach \i in {1,2,3,4,5}
  \node[ssnode,on chain] (t\i) [label=above right: {\small$\tau_\i$}] {};
\end{scope}

\begin{scope}[xshift=6cm,yshift=0cm,start chain=going below,node distance=7mm]
\foreach \i in {1,2,3,4,5}
  \node[ssnode,on chain] (tt\i) [label=above left: {\small$\overline\tau_\i$}] {};
\end{scope}

\begin{scope}[xshift=9cm,yshift=0cm,start chain=going below,node distance=7mm]
\foreach \i in {1,2,3}
  \node[ssnode,on chain] (c\i) [label=above right: {\small$\overline\gamma_\i$}] {};
\end{scope}

\node [circle, fill = black, inner sep = 0pt, minimum size = 5pt, label=left: $\xi$] at (-2, -0.9) (source) {};

\node [circle, fill = black, inner sep = 0pt, minimum size = 5pt, label=right: $\overline\xi$ ] at (11, -0.9) (sink) {};



\foreach \i in {1,2,3} {
   \draw[color=gray] (source) -- (r\i);
   }

\foreach \i in {1,2,3} {
   \foreach \j in {1,2,3,4,5} {
   \draw[color=gray] (r\i) -- (t\j);
   }
 } 
 
\foreach \i in {1,2,3,4,5} {
   \draw[color=gray] (t\i) -- (tt\i);
}  
 
\foreach \i in {1,2,3,4,5} {
   \foreach \j in {1,2,3} {
   \draw[color=gray] (tt\i) -- (c\j);
   }
} 

\foreach \i in {1,2,3} {
   \draw[color=gray] (c\i) -- (sink);
}

\node[text width=2cm, align = center] at (0,1.5) {\small\textsf{Top-30\% \\ articles}};
\node[text width=2cm, align = center] at (3,1.5) {\small\textsf{Bottom-70\% \\ articles}};
\node[text width=2cm, align = center] at (6,1.5) {\small\textsf{Mirror \\bottom-70\% \\ articles}};
\node[text width=2cm, align = center] at (9,1.5) {\small\textsf{Mirror \\ top-30\% \\ articles}};

\end{tikzpicture}
\caption{An illustrative plot of a tripartite network consisting of three treated units and five candidate control units.}
\label{fig: study design network}
\end{figure}
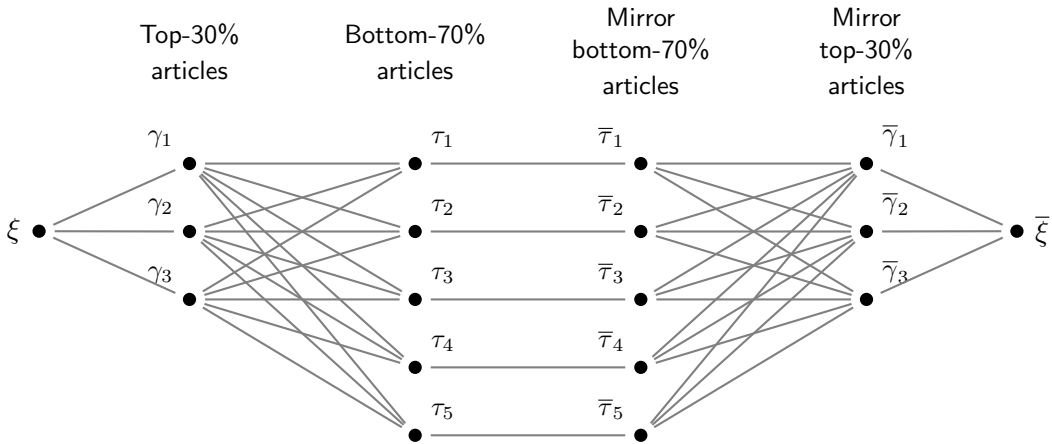

Figure \ref{fig: study design network} illustrates the basic tripartite network structure with three units, $\{\gamma_1, \gamma_2, \gamma_3\}$, from the top-30\% articles and five units, $\{\tau_1, \dots, \tau_5\}$, from the reservoir of bottom-70\% articles. To run a tripartite-network-based matching algorithm, two costs need to be specified \citep{zhang2021matching}. The \emph{cost} $\delta_{\gamma_i, \tau_j}$ associated with each edge $e$ connecting $\gamma_i$ and $\tau_j$ in the left part of the network encodes criteria for closely matched pairs. To construct the matched sample $\textsf{M1}$, we let $\delta_{\gamma_i, \tau_j}$ be the cosine similarity derived from the SPECTER embeddings of article $\gamma_i$ and $\tau_j$. We then enforce near-exact matching on submission year and reviewers' ratings by adding a large penalty to $\delta_{\gamma_i, \tau_j}$ if articles $\gamma_i$ and $\tau_j$ were submitted to the ICLR conference in different calendar years or did not receive same ratings. As a result, two articles in each matched pair were exactly or near-exactly matched on reviewers' ratings. For instance, a top-30\% article $\gamma_i$ submitted at $2017$ and peer-reviewed by three reviewers and received ratings of $6$, $5$ and $5$ (from highest to lowest) was matched to a $2017$-submitted, $(6, 5, 5)$-rated bottom-70\% article $\tau_j$. This conscious design of $\delta_{\gamma_i, \tau_j}$ helped achieve improved within-matched-pair cosine similarity and near-exact match on year of submission and reviewers' ratings in \textsf{M1}.

The cost $\Delta_{\overline\gamma_i, \overline\tau_j}$ associated with edge connecting $\overline{\gamma}_i$ and $\overline{\tau}_j$ in the right part of the network encodes criteria for good overall balance when groups are viewed as a whole. To construct $\textsf{M1}$, we estimated the propensity score based on article metadata and set $\Delta_{\overline\gamma_i, \overline\tau_j}$ to be the difference in the estimated propensity score to minimize earth-mover's distance between the propensity score distributions of the top-30\% articles and their matched comparison articles. The ``balancing" property of the propensity score \citep{rosenbaum1983central} then helped balance the covariates used to estimate it. One limitation of the propensity score is that its stochastic balancing property often does not suffice when balancing nominal variables with many categories; in these scenarios, a design technique known as \emph{fine balance} is often used in conjunction with propensity score matching \citep{rosenbaum2007minimum,rosenbaum2010design}. In short, fine balance is a technique that forces the frequency of one or more nominal variables to be identical or as close as possible in two groups. We finely balanced the topic clusters and keywords by adding a large penalty to $\Delta_{\overline\gamma_i, \overline\tau_j}$ when $\gamma_i$ and $\tau_j$ differed in the topic cluster or keyword. Finally, matched pairs in $\textsf{M1}$ were obtained as a result of solving the minimum-cost flow optimization problem associated with this tripartite network. We conducted matching in the stratum of articles with three reviewers and four reviewers, respectively, because articles in the four-reviewer stratum have two additional covariates: a fourth reviewer rating and a fourth reviewer sentiment.

Our construction of the design \textsf{M2} was analogous to that of \textsf{M1}, except that we further added a ``dose-caliper," defined as a large penalty when two articles $\gamma_i$ and $\tau_j$ differ in their authors' average institution rankings by less than a pre-specified caliper size, to the cost $\delta_{\gamma_i, \tau_j}$. The average within-matched-pair difference in authors' average rankings is $74.3$ in \textsf{M1}; hence, we set the caliper size to be $80$ when constructing \textsf{M2}. In this way, the within-matched-pair difference in authors' average ranking was as large as $108.8$ in the design \textsf{M2}, representing a meaningful improvement over that in \textsf{M1}.

\section{Discussion: interpretation of our results; limitations}
\label{sec: discussion}
In this article, we studied the association between author metadata and area chairs' decisions. We did \emph{not} find evidence supporting a \emph{status bias}, that is, area chairs' decisions systematically favored authors from high-ranking institutions, when comparing two cohorts of borderline articles with near-identical reviewers' ratings, sentiment, topics, primary keywords and article metadata. Under an additional stability assumption, we found that articles from high-ranking institutions had a lower acceptance rate and the result was consistent among articles from top-30\% institutions (odds ratio = $0.82$) and top-20\% institutions (odds ratio = $0.83$).  

Our results need to be interpreted under an appropriate context. First, like all retrospective, observational studies, although we formulated the question under a rigorous causal framework, we cannot be certain that our analysis was immune from any unmeasured confounding bias. For instance, the marginally higher acceptance rate of articles in the matched comparison groups (i.e., articles from lower-ranking institutions) could be easily explained away if these articles, despite having near-identical reviewers' ratings as their counterparts in the top-30\% or top-20\% groups, were in fact superior in their novelty and significance and area chairs made decisions based on these attributes rather than author metadata.

Second, any interpretation of our results should be restricted to our designed matched sample and should \emph{not} be generalized to other contexts. In particular, we only focused on area chairs' decision on \emph{borderline} articles. As \citet[Section III]{greiner2011causal} articulated in great detail, a study unit may interact with multiple agents of an overall system and we have more than one choice of decider to study. In our case study, an ICLR article has interacted with at least two types of deciders, a group of reviewers and an area chair. We explicitly focused on the interaction between a peer-reviewed article and an area chair. This deliberate choice allowed us to control for some valuable pre-treatment variables, including reviewers' ratings and sentiment, that are good proxies for articles' innate quality; however, by choosing to focus on this interaction that happened after the interaction between an article and multiple reviewers, we forwent the opportunity to detect any status bias, in either direction, in any earlier interaction including the peer review process that could have affected the values of pre-treatment variables in our analysis \citep{greiner2011causal}. Although the peer-review process of ICLR is in principle double-blind, it is conceivable that articles' author metadata could be leaked (e.g., when articles were posted in the pre-print platform) during the peer review process, and reviewers' ratings could be biased in favor of more (or less) established authors. It is of great interest to further study any perception bias during the interaction between articles and their reviewers; however, a key complication facing such an analysis is that articles may not be comparable without having a relatively objective judgment or rating (e.g., reviewers' ratings in our analysis of area chairs' decision).

With all these important caveats and limitations in mind, we found our analysis a solid contribution to the social science literature on status bias. Our analysis also helps clarify many important causal inference concepts and misconceptions when study units are textual data, including (1) the importance of shifting focus from an attribute to the perception of it, (2) the importance of articulating the timing of the treatment and hence what constitutes pre-treatment variables, (3) Fisher's sharp null hypothesis as a relevant causal null hypothesis in the context of fairness, and (4) \citeauthor{rubin1980discussion}'s \citeyearpar{rubin1980discussion} stability assumption often implicitly assumed but overlooked, all within a concrete case study.
\newcommand{\grantsponsor}[3]{#2}
\newcommand{\grantnum}[2]{#2}
\section*{Acknowledgements}
The authors would like to thank Weijie J.~Su
from the University of Pennsylvania
for stimulating discussions and helpful suggestions.
This work is in part supported by the
\grantsponsor{DARPA}{US Defense Advanced Research Projects Agency (DARPA)}{https://www.darpa.mil/}
under Contract \grantnum{DARPA}{FA8750-19-2-1004}; 
the \grantsponsor{ODNI}{Oﬃce of the Director of National Intelligence (ODNI)}{https://www.dni.gov/}, \grantsponsor{IARPA}{Intelligence Advanced Research Projects Activity (IARPA)}{https://www.iarpa.gov/}, via IARPA Contract No.~\grantnum{IARPA}{2019-19051600006} under the BETTER Program;
\grantsponsor{ONR}{Office of Naval Research (ONR)}{https://www.nre.navy.mil/} Contract \grantnum{ONR}{N00014-19-1-2620}; 
\grantsponsor{NSF}{National Science Foundation (NSF)}{https://www.nsf.gov/} under
Contract \grantnum{NSF}{CCF-1934876}.
The views and conclusions contained herein are those of the authors and should not be interpreted as necessarily representing the oﬃcial policies, either expressed or implied, of ODNI, IARPA, the Department of Defense, or the U.S.~Government. The U.S.~Government is authorized to reproduce and distribute reprints for governmental purposes notwithstanding any copyright annotation therein.

\section*{Appendix}
\label{sec: appendix}

\begin{figure}[H]
    \centering
    \includegraphics[width=\textwidth]{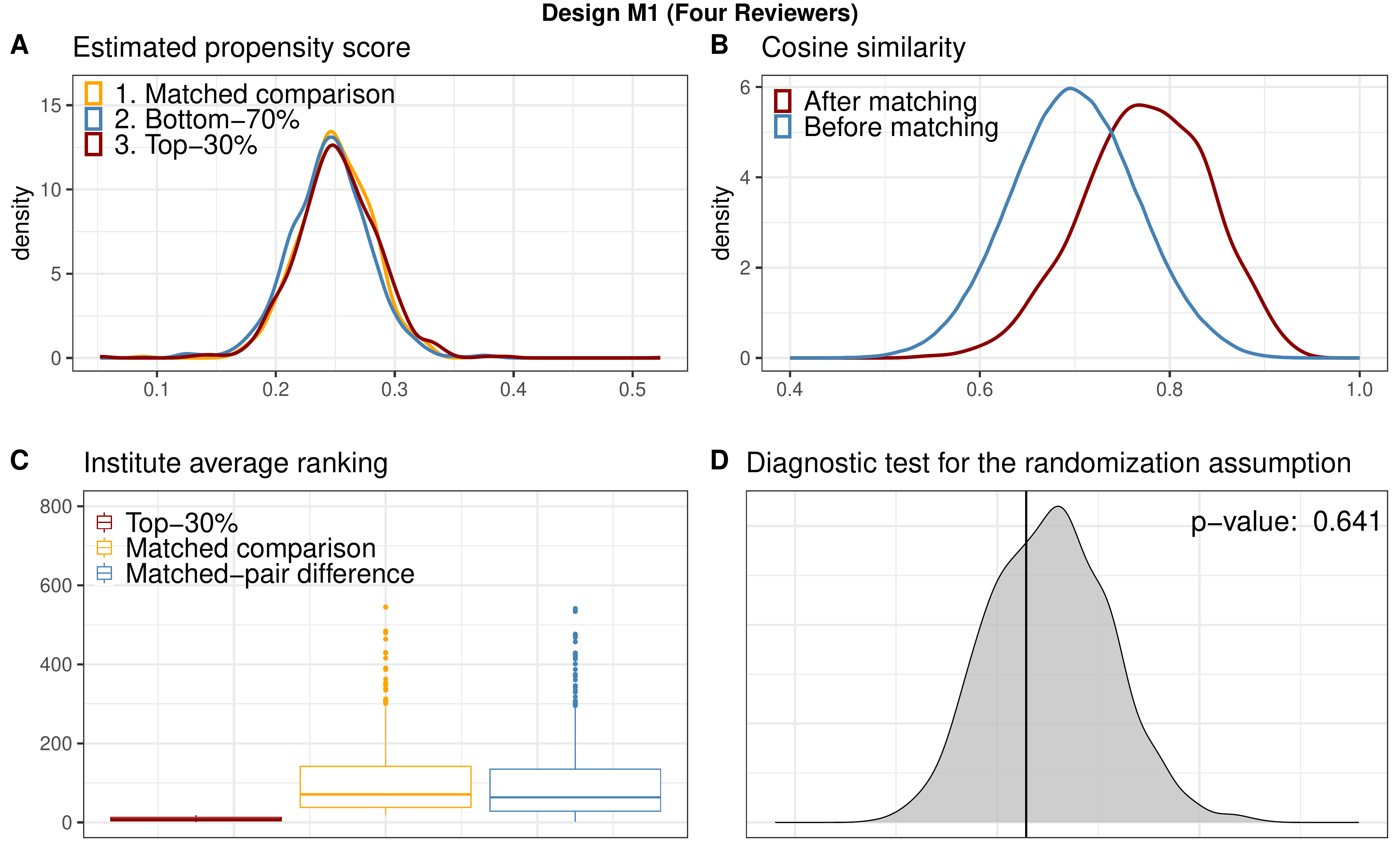}
    \caption{\small \textbf{Diagnostics of design $\textsf{M1}$}. Panel A: Estimated propensity score top-30\% articles, bottom-70\% articles and matched comparison articles. Panel B: Distribution of between-article cosine similarity before (median = 0.70) and after matching (median = 0.78). Panel C: Boxplots of authors' average institution rankings and matched-pair differences in authors' average institution rankings. The average, within-matched-pair difference in the institution ranking is 99.6 among 602 matched pairs in this stratum (median: 63.5; interquartile range 28.5-135.0). Panel D: Permutation distribution and the observed test statistic of the classification permutation test.}
    \label{fig: 30 4 reviewers combined}
\end{figure}

\begin{figure}[H]
    \centering
    \includegraphics[width=\textwidth]{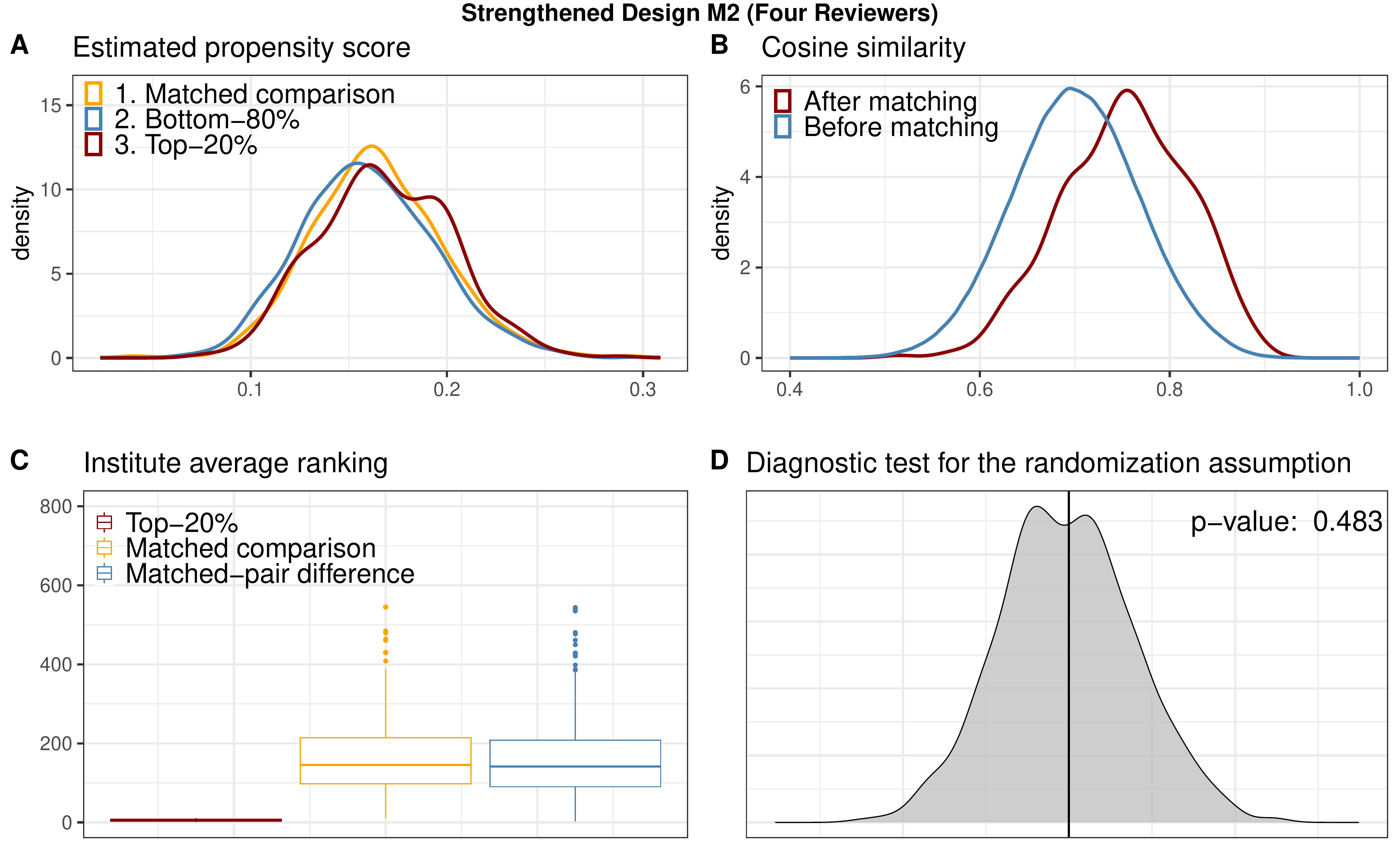}
   \caption{\small \textbf{Diagnostics of design $\textsf{M2}$}. Panel A: Estimated propensity score top-30\% articles, bottom-70\% articles and matched comparison articles. Panel B: Distribution of between-article cosine similarity before (median = 0.70) and after matching (median = 0.76). Panel C: Boxplots of authors' average institution rankings and matched-pair differences in authors' average institution rankings. The average, within-matched-pair difference in the institution ranking is 162.5 among 393 matched pairs in this stratum (median: 144.6; interquartile range 90.2-208.0). Panel D: Permutation distribution and the observed test statistic of the classification permutation test. }
    \label{fig: 20 4 reviewers combined}
\end{figure}

\clearpage
\bibliographystyle{apalike}
\bibliography{paper-ref,nlp-ref}

\end{document}